\documentclass[showkeys,prd,twocolumn,
]{revtex4-1}

\usepackage{hyperref}

\usepackage{graphicx}
\usepackage{dcolumn}
\usepackage{bm}
\usepackage{color}
\usepackage{epstopdf}
\usepackage{gensymb}
\usepackage[english]{babel}
\usepackage{xr}
\usepackage{sidecap}
\usepackage{mathrsfs,amsmath}
\usepackage[normalem]{ulem}

\newcommand{\bmat} {\begin{pmatrix}}
\newcommand{\emat} {\end{pmatrix}}

\begin{document}


\title{Acoustic Su-Schrieffer-Heeger lattice: \\
A direct mapping of acoustic waveguides to the Su-Schrieffer-Heeger model}
%
\author{Antonin Coutant}%
\affiliation{LAUM, UMR-CNRS 6613, Le Mans Universit\'e, Av. O. Messiaen, 72085 Le Mans, France}
\affiliation{Institut de Math\' ematiques de Bourgogne (IMB), UMR 5584, CNRS, Universit\' e de Bourgogne Franche-Comt\' e, F-21000 Dijon, France}
\email{antonin.coutant@univ-lemans.fr}
\author{Audrey Sivadon}
\affiliation{LAUM, UMR-CNRS 6613, Le Mans Universit\'e, Av. O. Messiaen, 72085 Le Mans, France}
\affiliation{LabTAU, INSERM, Centre L\' eon B\' erard, Universit\' e Lyon 1, Universit\' e Lyon, F-69003, LYON, France}
\author{Liyang Zheng} 
\affiliation{LAUM, UMR-CNRS 6613, Le Mans Universit\'e, Av. O. Messiaen, 72085 Le Mans, France}
\affiliation{Department of Physics, Universidad Carlos III de Madrid, ES-28916 Leganes, Madrid, Spain}
\author{Vassos Achilleos}
\affiliation{LAUM, UMR-CNRS 6613, Le Mans Universit\'e, Av. O. Messiaen, 72085 Le Mans, France}
\author{Olivier Richoux}
\affiliation{LAUM, UMR-CNRS 6613, Le Mans Universit\'e, Av. O. Messiaen, 72085 Le Mans, France}
\author{Georgios Theocharis}
\affiliation{LAUM, UMR-CNRS 6613, Le Mans Universit\'e, Av. O. Messiaen, 72085 Le Mans, France}
\author{Vincent Pagneux}
\affiliation{LAUM, UMR-CNRS 6613, Le Mans Universit\'e, Av. O. Messiaen, 72085 Le Mans, France}



\date{\today}
\begin{abstract}
Topological band theory strongly relies on prototypical lattice models with particular symmetries. We report here on a theoretical and experimental work on acoustic waveguides that are directly mapped to the one-dimensional Su-Schrieffer-Heeger model. Starting from the continuous two dimensional wave equation, we use a combination of 
monomode approximation and the condition of equal length tube segments to arrive at the wanted chiral symmetric discrete equations.
It is shown that open or closed boundary conditions leads automatically to the existence of topological edge modes. 
We illustrate by graphical construction how the edge modes appear naturally owing to a quarter-wavelength condition and the conservation of flux.
Furthermore, the transparent chirality of our system, which is ensured by simple geometrical constraints allows us to study chiral disorder numerically and experimentally. 
Our experimental results in the audible regime
demonstrate the predicted robustness of the topological edge modes.
%
%

\end{abstract}
\maketitle

\section{Introduction}

The field of topological insulators, first discovered in the context of the quantum Hall effect~\cite{Thouless82}, has found many applications to classical wave systems, whether it is in photonics~\cite{Ozawa19}, mechanics~\cite{Huber16}, or acoustics~\cite{Zhang18,Ma19} among others. A key aspect of materials/structures with topological phases is that they host modes localized on their boundaries or on designed interfaces, with properties that are robust to continuous deformations or the addition of special types of disorder. 

For one-dimensional systems, the most famous example of non-trivial topology is given by the Su-Schrieffer-Heeger (SSH) model~\cite{Su79,Asboth16}. 
This is the simplest two-band model where a topological phase transition occurs at band inversion coming with gap closing.
This discrete model belongs to the BDI class and as a consequence, in its topological phase, the system hosts localized modes on its edges. In particular, the chiral symmetry of the 1D SSH model guarantees that these edge modes have zero energy, a property that is maintained upon adding chiral disorder. 

The very appealing property of a localized mode with a locked disorder-insensitive frequency has generated a lot of interest in reproducing the 
behavior of the SSH model using  acoustic wave systems. Most of the approaches to date are based on coupled 
resonator systems~\cite{yang2016acoustic,Li18,esmann2018topological,shen2020acoustic,yan2020acoustic,chen2020chiral}.
There,  to mimic the SSH model,  the resonating acoustic cavities play the role of  the atoms  and the hoppings are realized by  connecting  tubes. 
The hopping strength is tunable by changing the cross-sectional areas of the tubes. 
The link with the SSH model is justified \emph{a posteriori} through a Tight Binding Approximation (TBA) with fitted coupling coefficients. 
Another approach  is based on waveguide phononic crystals~\cite{Xiao15,Meng18,yin2018band,Zangeneh19,Zangeneh20} where 
the analysis of the  band diagram owing to Zak phase enables the identifaction of topological properties of  band gaps.
In all these approaches, there is an analogy with the discrete SSH model.
However, the two main  ingredients of this model, the energy and the couplings, are not obtained through
explicit or simple expressions of the acoustic system. 

In this work, we report on an acoustic SSH lattice model based on a different approach. 
The idea relies on considering acoustic waveguides made of segments of alternating cross sections but more importantly, {\it equal lengths}~\cite{Dalmont94,Dalmont17,Zheng19,Zheng20,Coutant20,Coutant20b}. 
Starting from the 2D wave equation, for slender waveguide segments, we use a continuous monomode 1D approximation. 
Then, the choice of segments of equal length leads us to an explicit mapping with the 1D SSH model. 
On one hand, the SSH coupling coefficients are directly given by the ratios of cross-sections, and hence are easily tunable for numerical or experimental purposes, in contrast to all the previous acoustic approaches. On the other hand, the SSH energy appears naturally as a simple function of the acoustic frequency.

The monomode approximation used here, remains valid as long as the wavelength is large compared to the widths of the waveguide segments.
Consequently, in contrast to TBA,  our approximate discrete model is valid over a broad frequency range: from zero frequency up to the first cut-off frequency of the waveguide for which the 1D monomode approximation is broken. 
Surprisingly, the topological transition that accompanies the band inversion appears exactly for the case of the simplest of all waveguides, the one with constant cross section (this will be shown in Fig.~\ref{fig1_b}).
Another key consequence of our approach, and the explicit mapping to the 1D SSH model, is that chiral symmetry is ensured to be  preserved as long as the lengths of the segment are kept equal. This allows us to identify to what type of disorder the system is immune to.

The paper is organised as follows. In section~\ref{SSHmodel_Sec} we describe the acoustic waveguide and show its explicit mapping with the SSH model. In section~\ref{FiniteSSH_Sec} we compute the set of eigenmodes for a finite waveguide with different boundary conditions, and discuss the presence of topological edge and interface modes. The influence of disorder is analyzed in section~\ref{DisorderSSH_Sec}, and in section~\ref{Exp_Sec}, we present the experimental results.

\section{From 2D Helmholtz to SSH model}
\label{SSHmodel_Sec}

\begin{figure}[t!]
\includegraphics[width=\columnwidth]{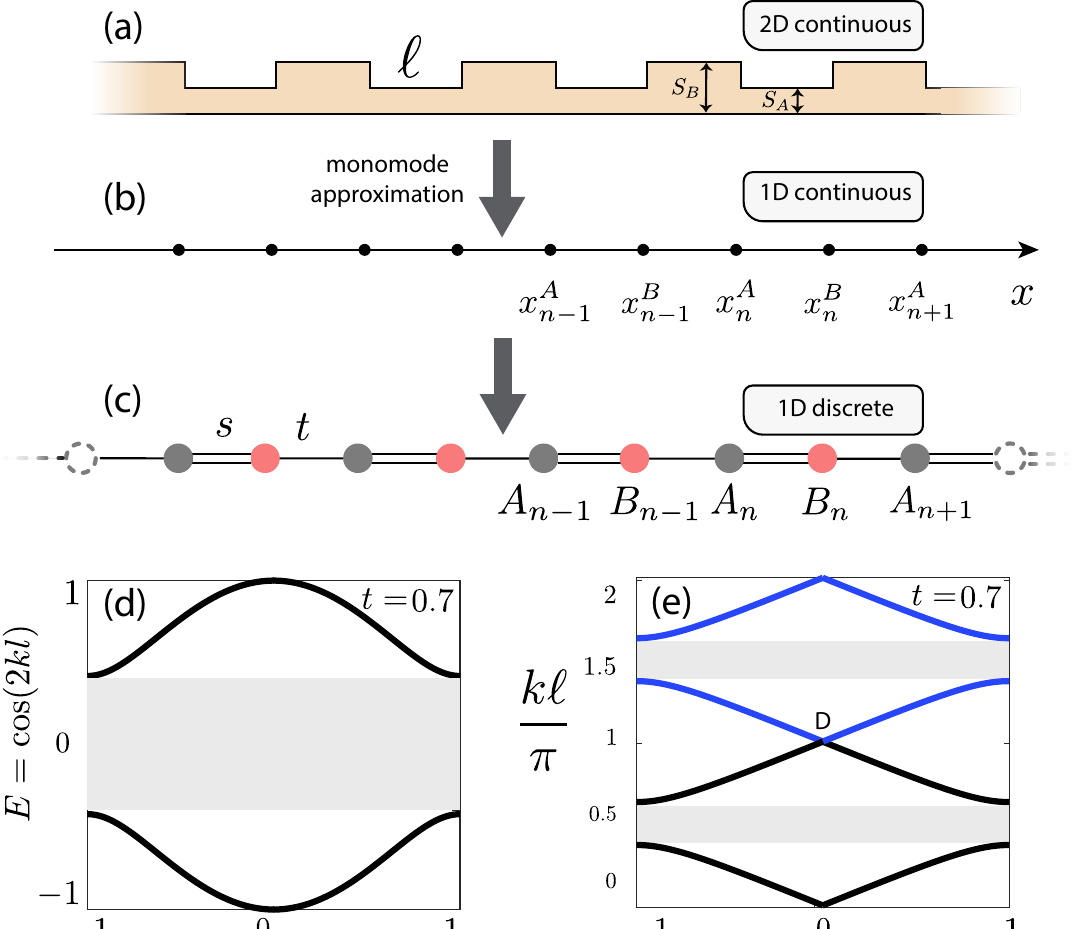}
\caption{(a) A sketch of a 2D acoustic waveguide composed of segments of the same length $\ell$ and different cross sections
$S_A$ and $S_B$. (b) Under the monomode approximation the acoustic pressure depends only on the axial coordinate $x$. On this 
axis, we can identify the points where the cross-sections change. (c) The explicit map onto the SSH discrete model. (d)-(e) The dispersion relation of the SSH model as given by equation~(\ref{disp}).
}
\label{fig0} 
\end{figure}

We consider an acoustic waveguide, as shown in Fig.~\ref{fig0}(a), composed of periodically arranged segments of length $\ell$ with two different cross sections $S_A$ and $S_B$.  
For a lossless fluid  in the linear regime with time harmonic dependence $e^{-i \omega t}$, the  acoustic pressure field $p(x,y)$ is governed by the 2D Helmholtz equation
\begin{equation}
\frac{\partial ^2 p}{\partial x^2}+\frac{\partial ^2 p}{\partial y^2}+ k^2 p = 0 , \label{eq_H2}
\end{equation}
with Neumann boundary conditions, $\partial_n p =0$ at the boundaries, corresponding to zero normal velocity  at the rigid wall [solid black lines in Fig1. (a)].
Here $k=\omega/c$  with $\omega$ the angular frequency and $c$ the sound velocity.

For sufficiently low frequencies, when only one mode is propagating in each waveguide (monomode assumption), it is possible to make the approximation that, in each segment, the acoustic wave is governed  by the one-dimensional (1D) 
wave  equation
\begin{equation}
\frac{d^2 p}{dx^2} + k^2 p = 0 , \label{eq_prop}
\end{equation}
where the pressure $p(x)$ depends only the axial coordinate $x$. 
The most important part in this simple 1D approximation is in the jump  conditions at each cross-section $S$ change:
\begin{align}
[p] =0, \; \; [S \frac{d p}{dx}] &=0, \label{eq_BC2}
\end{align}
 enforcing  actually continuity of acoustic pressure and flow rate~\cite{Dalmont94,Dalmont17}. Here $[X]=X^+-X^-$ denotes the difference at each cross-section.  
At this stage, we have reduced the initial 2D continuous problem (Fig.~\ref{fig0}(a)) to a 1D continuous approximation (Fig.~\ref{fig0}(b)).

Now, we can derive the corresponding discrete equations by focusing only on the acoustic pressure at the points of cross-section area changes.
From equation~\eqref{eq_prop}, the trick is to write the following 
expressions  corresponding to the solutions of equation~\eqref{eq_prop} for the segments
between $x_n^A, x_n^B$ and  $x^B_{n-1} , x_n^A$ respectively	
\begin{align}
p(x_n^{B}) &= \cos{k\ell} \; p(x_n^{A})+\frac{\sin{k\ell}}{k} p' (x_n^{A+}),\label{eq_t1}\\
p(x_{n-1}^{B}) &= \cos{k\ell} \; p(x_n^{A})-\frac{\sin{k\ell}}{k} p' (x_n^{A-}) \label{eq_t2}
\end{align}
where prime denote $d/dx$. Here, the continuity of pressure Eq.~\eqref{eq_BC2} has been already used, leading to no differentiation between pressure at left (-) and right (+). 
The pressure derivatives are discontinuous at the cross-section changes, but
we remark that they can be eliminated using  the relations given by equations~(\ref{eq_BC2}); 
indeed multiplying equation~\eqref{eq_t1} by $S_A$, equation~\eqref{eq_t2} by $S_B$ and summing yields
\begin{equation}
S_A \; p(x_n^{B}) + S_B \; p(x_{n-1}^{B}) = (S_A +S_B) \cos{k\ell} \;  p(x_n^{A}). \label{eq_p1}
\end{equation}
Following directly the same lines, but around $x=x_n^{B}$, we  obtain 
\begin{equation}
S_A \; p(x_n^{A}) + S_B \; p(x_{n+1}^{A}) = (S_A +S_B) \cos{k\ell} \;  p(x_n^{B}). \label{eq_p2}
\end{equation}
Eventually, the entire  periodic waveguide is described by the following equations connecting acoustic pressure at  consecutive changes of section as 
\begin{align}
 s B_n+ t B_{n-1}  = E(k) A_n,\label{eq1} \\
t A_{n+1} + s A_n   = E(k) B_n, \label{eq2} 
\end{align}
where $A_n \equiv p(x_n^A)$  and $B_n \equiv p(x_n^B)$, 
\begin{align}
s = \frac{S_A}{S_A+S_B} \quad ,  \quad
t = \frac{S_B}{S_A+S_B} . \label{teq} 
\end{align}
and
\begin{equation}
E(k) = \cos k \ell . \label{eqE} 
\end{equation}
For the infinite system, it corresponds to the eigenvalue problem $H {\bf X} = E  {\bf X} $ with
\begin{equation}
H = \bmat \ddots & \ddots & &  & &  \\ 
\ddots & 0 & s & & &  \\ 
 & s & 0 & t & &  \\ 
 & & t &0 & s &  \\ 
& & & s & 0 & \ddots \\
 &  & & & \ddots & \ddots \emat,
 \label{hami}
\end{equation}
and the vector ${\bf X}$ containing the unknown pressure values $A_n$ and $B_n$, ${\bf X} = \bmat \dots, A_n, B_n, \dots \emat^T$. 
Note that here for our acoustic problem, the coupling coefficients are positive, smaller than unity and satisfy the relation (see equation~\eqref{teq}): 
\begin{equation}
s+t=1
\end{equation}
It appears that the set of equations \eqref{eq1}-\eqref{eq2} coincides exactly to the SSH model, which is represented in Fig.~\ref{fig0}(c). 
We note that in the proposed acoustic  waveguide, the hopping coefficients $s$ and $t$ of the SSH model,  
equation \eqref{teq}, are  given directly by the geometrical cross-section ratios.
This explicit and simple expression of the hoppings has to be contrasted with the one involved in the  classical tight binding approximation that
requires local resonators (with the associated \emph{a priori} narrow band validity) and fitting of hopping coefficients.
Furthermore in our approximate 1D modeling, the acoustic pseudo-energy $E(k) = \cos k \ell $ is the explicit analogue of the energy of the SSH model.
It means that the eigenfrequencies $k$ are obtained directly from the eigenvalues E of the SSH Hamiltonian $H$ from equation~\eqref{hami}.
All we need is the 1D approximation to remain valid, which is true as long as
(i) the aspect ratio of the segments are small enough 
(ii) the wavelength is large compared to the width of the waveguide. This 1D approximation is broadband in the sense that it is valid from zero frequency to some upper bound where the wavelength is no longer respecting condition (ii).
Consequently, the frequency range (from zero to the upper band frequency) can be as large as wanted by choosing thin enough segments. Of course, when going to extremely thin segments, taking care of the visco-thermal losses will be necessary. In the sequel, we consider segments of moderate aspect ratio (typically of the order of 0.2) which are easily constructed, exhibit minor losses and where the 1D approximation is fairly accurate. We notice here that if the tubes have different lengths, one cannot eliminate pressure derivatives from equations~\eqref{eq_t1}, \eqref{eq_t2} to equations \eqref{eq_p1}, \eqref{eq_p2}, and hence, the mapping with the SSH lattice model does not hold. As a result, the edge modes that will be studied in the sequel will lose their 
robustness to random changes (in some sense, different tube lengths introduce a breaking of the chiral symmetry).
\begin{figure*}[!ht]
\includegraphics[width=0.9\textwidth]{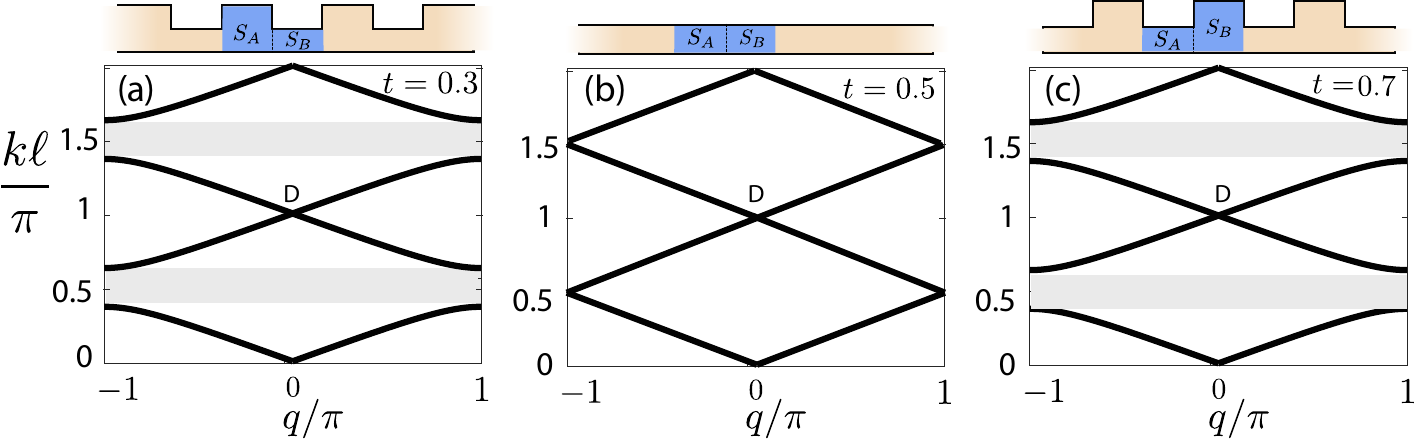}
\caption{The dispersion relation of the acoustic SSH for three different values of the coupling  $t=0.3$ (a), $t=0.5$ (b), $t=0.7$ (c).
This illustrates the band inversion of the model. The crossing of the two bands of the SSH model at $s=t$ appears exactly for the simple waveguide with
constant cross section as shown in panel (b).}
\label{fig1_b} 
\end{figure*}

Now we follow what is offered by the SSH model. 
For the case of infinite periodic waveguide, we may assume Bloch wave solution given by $A_n = A e^{iqn}$ and $B_n = B e^{iqn}$ where $q$ is the Bloch wavenumber and from equations~(\ref{eq1},\ref{eq2}) we obtain the following eigenvalue problem 
\begin{equation}  \label{eq_eigen}
\left(
\begin{matrix}
0 & s+t e^{-iq}  \\
s+t e^{iq} & 0 
\end{matrix} 
\right)
\left( \begin{matrix}
A \\
B
\end{matrix}
\right)
=
E(k)
\left(\begin{matrix}
A \\
B
\end{matrix}
\right).
\end{equation}
Equation (13) immediately  shows that, using the proposed simple system we  recovered the chiral $2\times 2$ Hamiltonian matrix of the periodic 1D SSH model  where the coupling coefficients are simply given by the ratios of the two different cross sections. 
The dispersion relation can be found from equation~\eqref{eq_eigen} as
\begin{equation}
E(k) = \cos k\ell = \pm \sqrt{s^2+t^2+2st \cos q}.
\label{disp}
\end{equation}
This dispersion relation, emblematic of the SSH model, is shown in Fig.~\ref{fig0}(d) as a function of the energy $E$, 
while in Fig.~\ref{fig0}(e) we plot the dispersion relation in terms of the acoustic wavenumber $k\ell/\pi$.

Due to chiral symmetry~\cite{Asboth16}, the spectrum is symmetric around $E(k) = 0$ which corresponds to $k \ell = (n+1/2) \pi$. 
Here it is interesting to illustrate, for our acoustic system, the corresponding band inversion process. This is shown in Fig.~\ref{fig1_b}, where the waveguide spectrum is shown for three different values of the cross-section ratio $t$. The most intriguing aspect of the band inversion occurring at $t=1/2$ is that it corresponds to $S_A=S_B$, which means a flat waveguide as shown in Fig.~\ref{fig1_b} (b). Let us stress here that there is no need to compute the topological Zak phase of the bands since our acoustic waveguide is modelled by the discrete SSH system where the
topological transition is classically obtained from the winding number~\cite{Asboth16}.

\section{Finite system}
\label{FiniteSSH_Sec}
To further investigate the acoustic SSH model, in this section, we consider waveguides of finite size (finite number of segments), and study their eigenmodes. For this, we must solve an eigenvalue problem for a finite sized matrix (the Hamiltonian $H$), which size corresponds to the number of sites in the lattice model. We then compare the results with two-dimensional numerical simulations using a finite element method. 

Different types of Boundary Conditions (BC) can be considered, open with Dirichlet BC or hard wall with Neumann BC, and in the next sections we are 
looking at these different BC's for a finite waveguide system.

\subsection{Open boundary conditions and an even number of cross section changes}
\begin{figure}[t!]
\includegraphics[width=\columnwidth]{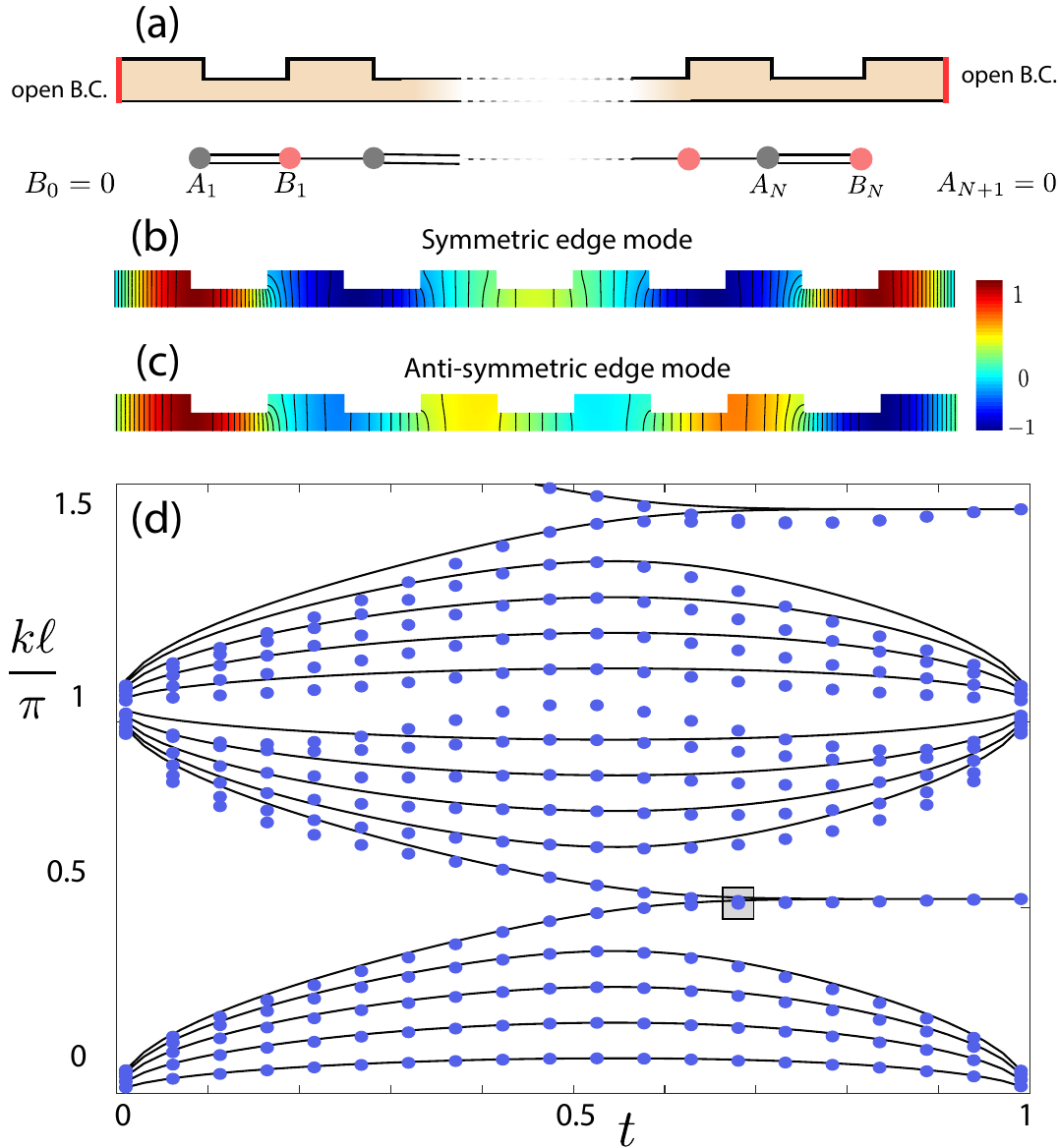}
\caption{Open boundary conditions and even cross section changes. (a) Schematic representation of the waveguide configuration with open boundary conditions and the lattice model corresponding to the one-dimensional limit of the waveguide. The symmetric (b) and anti-symmetric (c) edge modes as obtained numerically for a waveguide with $2N+1=11$ and $t=2/3$. (d) The spectrum of the same waveguide of $2N+1=11$ segments as a function of $t$.}
\label{fig1} 
\end{figure}

We first consider finite waveguides made of $2N+1$ segments and $2N$ cross section changes, where both ends have a cross section $S_B$ and are open to the exterior, as shown in Fig.~\ref{fig1}(a). Equilibrium with the exterior pressure imposes the boundary condition of vanishing pressure at these ends (we neglect radiative losses, which vanish in the limit of small cross sections). In the lattice representation, this boundary condition amounts to having a site on the left (resp. right) end that is only connected to a right (resp. left) neighbour, i.e. $B_0 = A_{N+1} = 0$. This is represented in Fig.~\ref{fig1}(a). We compute the discrete set of eigenmodes, which are the eigenvectors of the Hamiltonian
\begin{equation} \label{EvenOpen_H}
H = \bmat 0 & s & 0 & \dots & & 0 \\ 
s & 0 & t & & & \vdots \\ 
0 & t & 0 & \ddots & &  \\ 
\vdots & & \ddots & \ddots & t & 0 \\ 
& & & t & 0 & s \\
0 & \dots & & 0 & s & 0 \emat , 
\end{equation}
and the eigenvalue problem $E {\bf X} = H \cdot {\bf X}$ with ${\bf X} = \bmat A_1, B_1, \dots, A_N, B_N \emat^T$. 
Bulk modes can be obtained as superpositions of right and left going Bloch waves (i.e. eigenvectors of equation~\eqref{eq_eigen}). Assuming $E>0$, we look for modes of the form 
\begin{equation} \label{Bloch_superpo}
\bmat A_n \\ B_n \emat = \alpha e^{i q n} \bmat 1 \\ e^{i \phi(q)} \emat + \beta e^{-i q n} \bmat 1 \\ e^{-i \phi(q)} \emat , 
\end{equation}
where $\phi(q) = \arg(s + t e^{iq})$, and $\alpha$ and $\beta$ are complex amplitudes to be determined. Requiring that the end sites are only one-side connected (or equivalently, the pressure vanishes on the open ends) leads to the relation $\alpha = -\beta$, and the equation 
\begin{equation} \label{Delplace_eq}
\sin \left((N+1) q + \phi(q) \right) = 0, 
\end{equation}
which has to be solved for $0 < q < \pi$ to obtain bulk modes and their eigenenergies. Interestingly, the phase $\phi(q)$ is related to the so-called Zak phase, which characterize the topology of the SSH model. Based on this, equation \eqref{Delplace_eq} can be used to demonstrate the bulk-edge correspondence in this model~\cite{Delplace11} (see also~\cite{Coutant20b} for a higher order topology context). Each solution for the Bloch wavenumber $q$ gives an eigenenergy through the relation $E(q) = |s+t e^{iq}|$ (with its opposite $-E(q)$ being also an eigenenergy by chiral symmetry). 

We now compute the spectrum for varying cross sections using both the 1D discrete model (direct diagonalization of \eqref{EvenOpen_H}) and 2D numerical simulations (finite element method). They are shown in Fig.~\ref{fig1}(d), with a good agreement between the discrete and continuous models. We see two distinct cases: if $s>t$, all eigenvalues are concentrated in bands of the infinite model and correspond to bulk modes; if $s<t$ we see two eigenvalues inside the gaps. These isolated eigenvalues correspond to eigenmodes localized on the edges (symmetrically or anti-symmetrically), and are the topologically protected modes of the SSH model~\cite{Asboth16}. The hybridization of these two edge states, induces a small energy splitting near $t=1/2$. The 2D modes are shown in Figs.~\ref{fig1}(b,c). 

\subsection{Open boundary conditions and an odd number of cross section changes}
\begin{figure}[t!]
\includegraphics[width=\columnwidth]{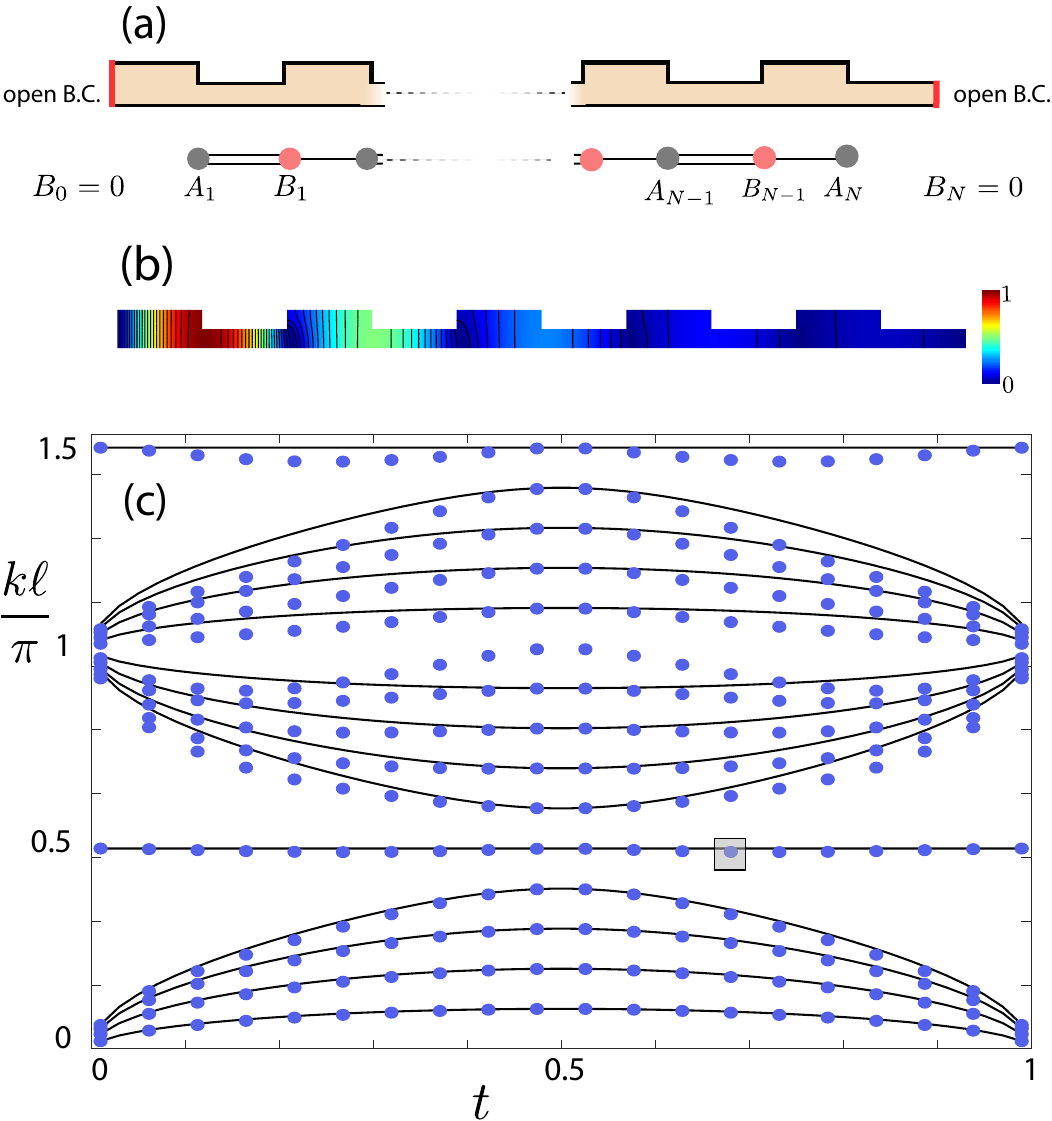}
\caption{Same as in  Fig.~\ref{fig1} but for the case of open boundary conditions and even cross section changes.
}
\label{fig2} 
\end{figure}

Next, we consider a modified configuration, which is obtained by taking out the last segment on the right hand side of the previous one~\cite{shen2020acoustic}. Doing so, the segments on the left end and right end have different cross section values. At the level of the lattice model, this amounts to having an odd number of sites ($2N-1$) and $B_0=B_N=0$. This new configuration is shown in Fig.~\ref{fig2}(a), and corresponds to the Hamiltonian
\begin{equation}
H = \bmat 0 & s & 0 & \dots & 0 \\ 
s & 0 & t & & \vdots \\ 
0 & t & 0 & \ddots & 0  \\ 
\vdots & & \ddots & \ddots & t \\ 
0 & \dots & 0 & t & 0 \emat , 
\end{equation}
and the eigenvalue problem $E {\bf X} = H \cdot {\bf X}$ with ${\bf X} = \bmat A_1, B_1, \dots, B_{N-1}, A_N \emat^T$. 
Bulk modes can again be obtained as superposition of Bloch waves as in equation~\eqref{Bloch_superpo}, but the equation for the extra site now leads to the simpler equation 
\begin{equation}
\sin \left((N+1) q\right) = 0, 
\end{equation}
which again has to be solved for $0 < q < \pi$. The main advantage of this new configuration is that there is always a unique edge mode: if $s>t$ it is localized on the right end, and if $s<t$ it is localized on the left end. The edge mode for $t=2/3$ is shown in Fig.~\ref{fig2}(b). In Fig.~\ref{fig2}(c), we show the spectrum for varying cross sections. The second advantage of this configuration is that the edge mode has a closed-form expression. Indeed, the eigenvalue problem can be written as $2N-1$ equations, which coincide with equations \eqref{eq1} and \eqref{eq2} for $n=1..N-1$. For $E=0$, the two equations are independent. Moreover, the boundary conditions require $B_0 = B_{N} = 0$, which implies $B_n = 0$ for all $n$. 
Equation \eqref{eq1} is straightforward to solve and leads to the edge mode 
\begin{equation} \label{Periodic_EdgeMode_OpenOdd}
\bmat A_n \\ B_n \emat = \alpha_0 \bmat 1 \\ 0 \emat \left(-s/t \right)^n , 
\end{equation}
where $\alpha_0$ is a normalization constant. This corresponds to $p(x_n^A) = \alpha_0 (-s/t)$ and $p(x_n^B) = 0$. 

Hence, this configuration displays an edge state for all parameter values. This could appear as a contradiction of the bulk boundary correspondence, since the topological invariant (Zak phase or winding number~\cite{Asboth16}) changes when passing through the Dirac point at $s=t$. We recall that this apparent paradox is resolved when one notes that the two edges in the SSH model with an odd number of sites correspond to different choices of unit cells~\cite{Prodan16}. Indeed, a given edge is compatible with a choice of unit cell if the chain ends on the edge with a whole unit cell (the same discussion extends to 2D systems such as honeycomb lattices~\cite{Delplace11,Kariyado17}). Moreover, different unit cells lead to different winding numbers. Hence, the bulk boundary correspondence is restored by looking at each boundary separately, and the winding number going with the compatible unit cell. In the present configuration, this calculation gives 1 edge mode on the left and 0 edge mode on the right when $s<t$, and the inverse for $s>t$. 

\subsection{Closed boundary conditions and an odd number cross section changes}
\begin{figure}[t!]
\includegraphics[width=\columnwidth]{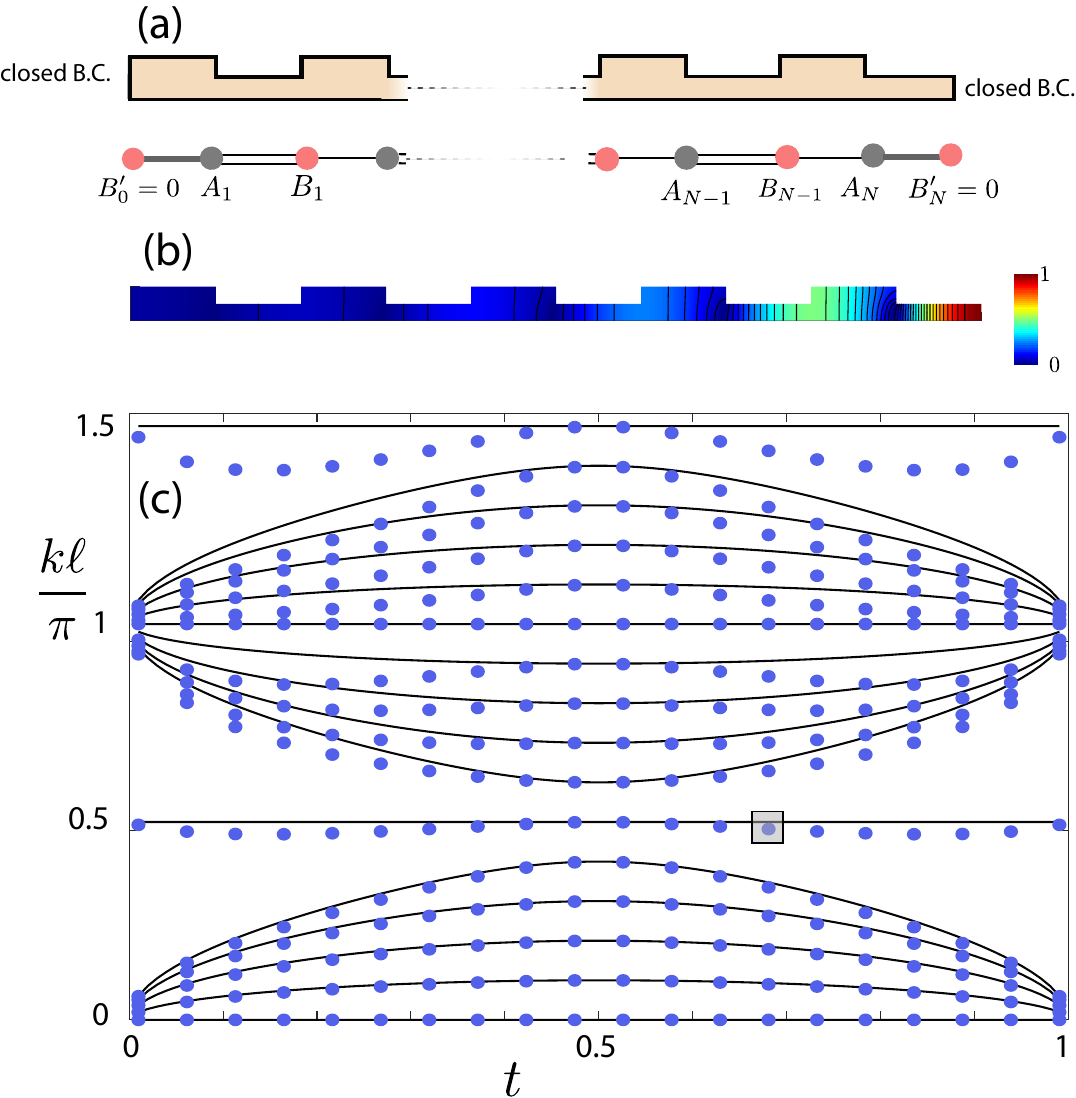}
\caption{Same as in  Fig.~\ref{fig1} but for the case of closed boundary conditions and even cross section changes. 
}
\label{fig3} 
\end{figure}

We now consider the same configuration as in the previous subsection, but with closed walls on both ends, hence changing the boundary conditions: the acoustic velocity now vanishes on both ends. For the lattice model, this introduces two particular sites on both sides. These sites obey a different equation due to the boundary condition. Using a procedure similar to section~\ref{SSHmodel_Sec}, we have that $E(k) B_0 = A_1$, and $E(k) B_{N+1} = A_N$. This leads to the Hamiltonian  
\begin{equation}
H = \bmat 0 & 1 & 0 & \dots & 0 \\ 
s & 0 & t & & \vdots \\ 
0 & t & 0 & \ddots & 0  \\ 
\vdots & & \ddots & \ddots & t \\ 
0 & \dots & 0 & 1 & 0 \emat ,
\end{equation}
and the eigenvalue problem $H\cdot {\bf X} = E {\bf X}$ with ${\bf X} = \bmat B_0, A_1, B_1, \dots, A_N, B_N \emat^T$. Note that the matrix $H$ is not hermitian, but it can be made so by a similarity transformation (we will address this in section~\ref{DisorderSSH_Sec}). 
This configuration, its spectrum and edge mode are shown in Fig.~\ref{fig3}. We see that the results are very similar to the previous configuration (see Fig.~\ref{fig2}). 
In view of the experimental implementation, this last configuration is preferred because open ends induce radiative losses while closed walls does not. Similarly to equation \eqref{Periodic_EdgeMode_OpenOdd}, the edge mode has a closed-form expression 
\begin{equation} \label{Periodic_EdgeMode_CloseOdd}
\bmat A_n \\ B_n \emat = \alpha_0 \bmat 0 \\ 1 \emat \left(-s/t \right)^{-n} . 
\end{equation}
Notice that the edge mode is now localized on the other side of the system, as is manifest from equation~\eqref{Periodic_EdgeMode_CloseOdd}. 

\subsection{Graphical construction of the edge mode}
\begin{figure}[t!]
\includegraphics[width=\columnwidth]{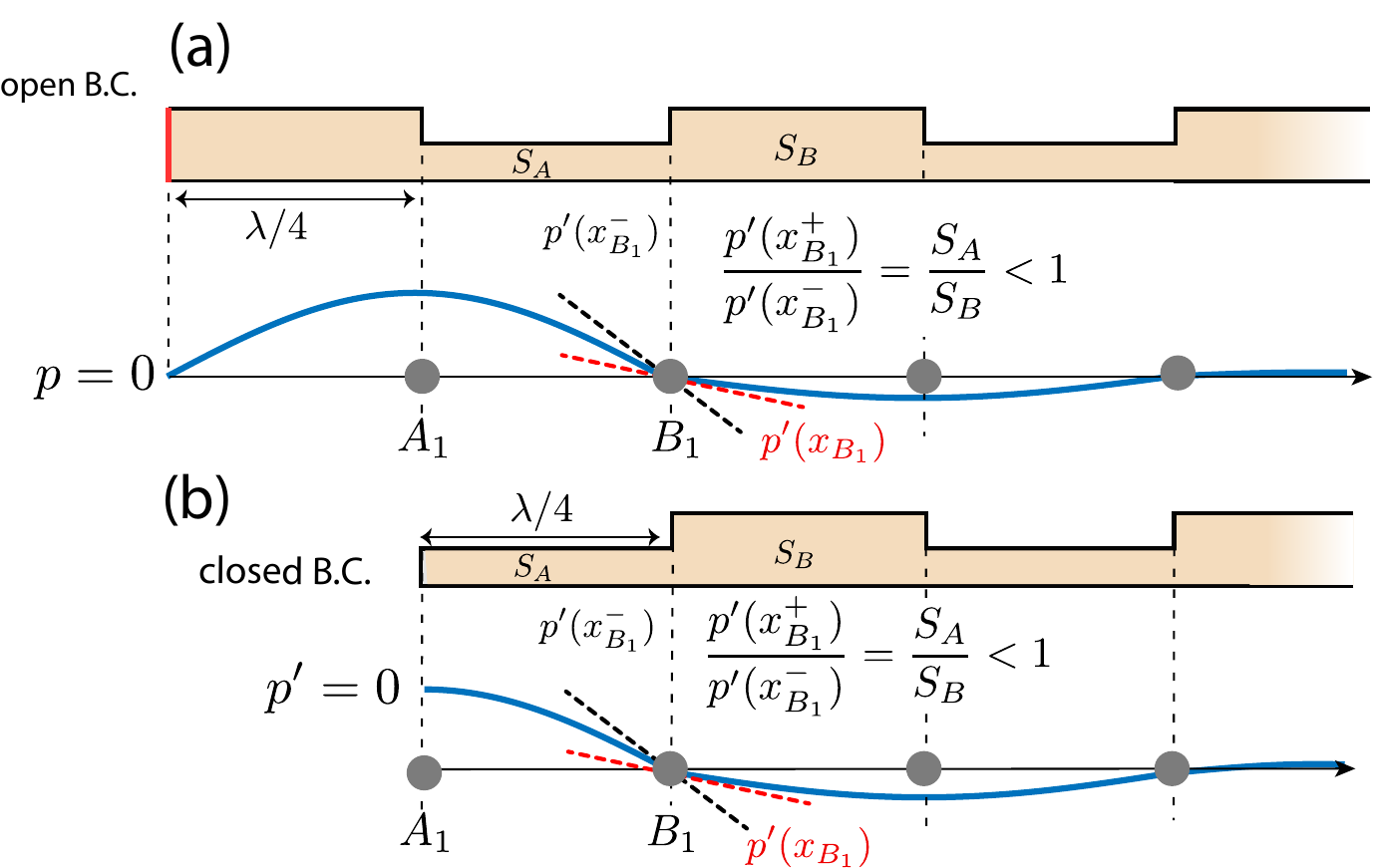}
\caption{Graphical construction of the edge modes corresponding to the case of open (a) and closed (b) boundary conditions. The dashed lines
at point $B_1$ illustrate the reduction of slope  and thus of the oscillation amplitude.}
\label{fig4} 
\end{figure}

In our acoustic waveguides, from the very definition of the pseudo energy equation~\eqref{eqE}, it appears that the first zero energy mode ($E=0$) of the SSH model corresponds to a quarter wavelength ($k\ell = \pi/2$ i.e. $\lambda/4 = \ell$) pressure field in each of the segment of length $\ell$.
Then, it is possible to have a very simple and hopefully enlightening explanation of this edge mode that is of graphical nature.
In Fig.~\ref{fig4}, we display this construction for both the case of open waveguide (zero pressure at the extremity in Fig.~\ref{fig4}(a)) and of
hard wall termination (zero derivative of the pressure in Fig.~\ref{fig4}(b)).

In Fig.~\ref{fig4}(a) for an open waveguide, we  see how the edge mode is able to comply with the succession of cross-section changes:
 starting from the edge boundary condition $p=0$, due to the $\lambda/4 = \ell$ relation, each decrease of cross-section (as at point $A_1$) is properly ignored (here $p'=0$ and the change is transparent) whilst each increase of cross-section (as at point $B_1$) induces a reduction of the slope and consequently a reduction of the oscillation amplitude. The mechanism of zero energy edge mode is the same for the hard wall case [Fig.~\ref{fig4}(b)] starting from the left extremity where $p'=0$. In this construction, once again, we can see that what is the important point to map to SSH model is the identical length of each of the waveguide segments.

\begin{figure}[t!]
\includegraphics[width=\columnwidth]{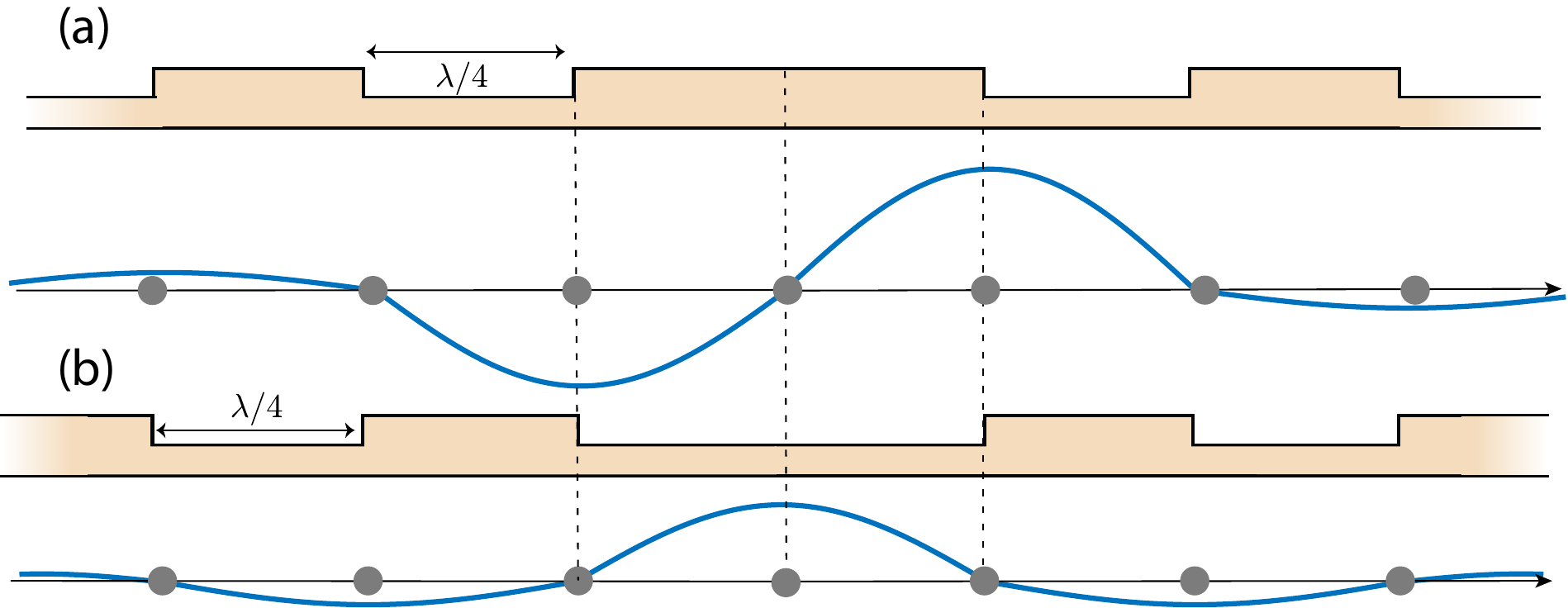}
\caption{Graphical construction of interface modes by connecting two semi-infinite periodic waveguides of different topologies.
(a) antisymmetric case obtained from Fig. \ref{fig4}(a). (b) symmetric case obtained from Fig. \ref{fig4}(b)}
\label{fig4b} 
\end{figure}

Incidentally, we can remark that interface modes obviously exist in our acoustic waveguide.
 They follow exactly the same graphical construction as seen in  Fig.~\ref{fig4b}. 
There are represented on 
the interfaces joining two semi-infinite periodic parts with different topology. We see that it corresponds simply to unfolding the two previous edge configurations 
Fig.~\ref{fig4b}(a) coming from Fig.~\ref{fig4}(a) and 
Fig.~\ref{fig4b}(b)  from Fig.~\ref{fig4}(b).  

\section{Disorder}
\label{DisorderSSH_Sec}

We now introduce disorder in the closed system by changing the cross section $S_m$ of each segment ($m\in[1;2N]$). We  keep the length $\ell$ constant in order to preserve the chiral
symmetry of the SSH model. Indeed, we can rederive the discrete equations as in section~\ref{SSHmodel_Sec}. Again, we exploit continuity of pressure and acoustic flow rate at each cross section change. Let us start by considering the pressure at the position $x=x_n^A$. As before, we integrate the 1D Helmholtz equation~\eqref{eq_prop} from $x_{n-1}^B$ to $x_n^A$ and then from $x_{n}^B$ to $x_n^A$, and eliminate pressure derivatives using acoustic flow rate conservation. We then do the same starting from $x=x_n^B$. We obtain the two sets of equations: 
\begin{eqnarray} \label{DisorderSSH_nonNorm1}
E(k) A_n &=& \frac{S_{2n-1}}{S_{2n-1}+S_{2n}} B_{n-1} + \frac{S_{2n}}{S_{2n-1}+S_{2n}} B_n , 
\end{eqnarray}
with $n \in[1;N]$, 
\begin{eqnarray} \label{DisorderSSH_nonNorm2}
E(k) B_n &=& \frac{S_{2n}}{S_{2n}+S_{2n+1}} A_{n} + \frac{S_{2n+1}}{S_{2n}+S_{2n+1}} A_{n+1} 
\end{eqnarray}
with $n \in[0;N]$. Notice that the closed boundary conditions are easily implemented, as they correspond to defining extra cross section values $S_{0} = S_{2N+1} = 0$. 

\begin{figure}[t!]
\includegraphics[width=\columnwidth]{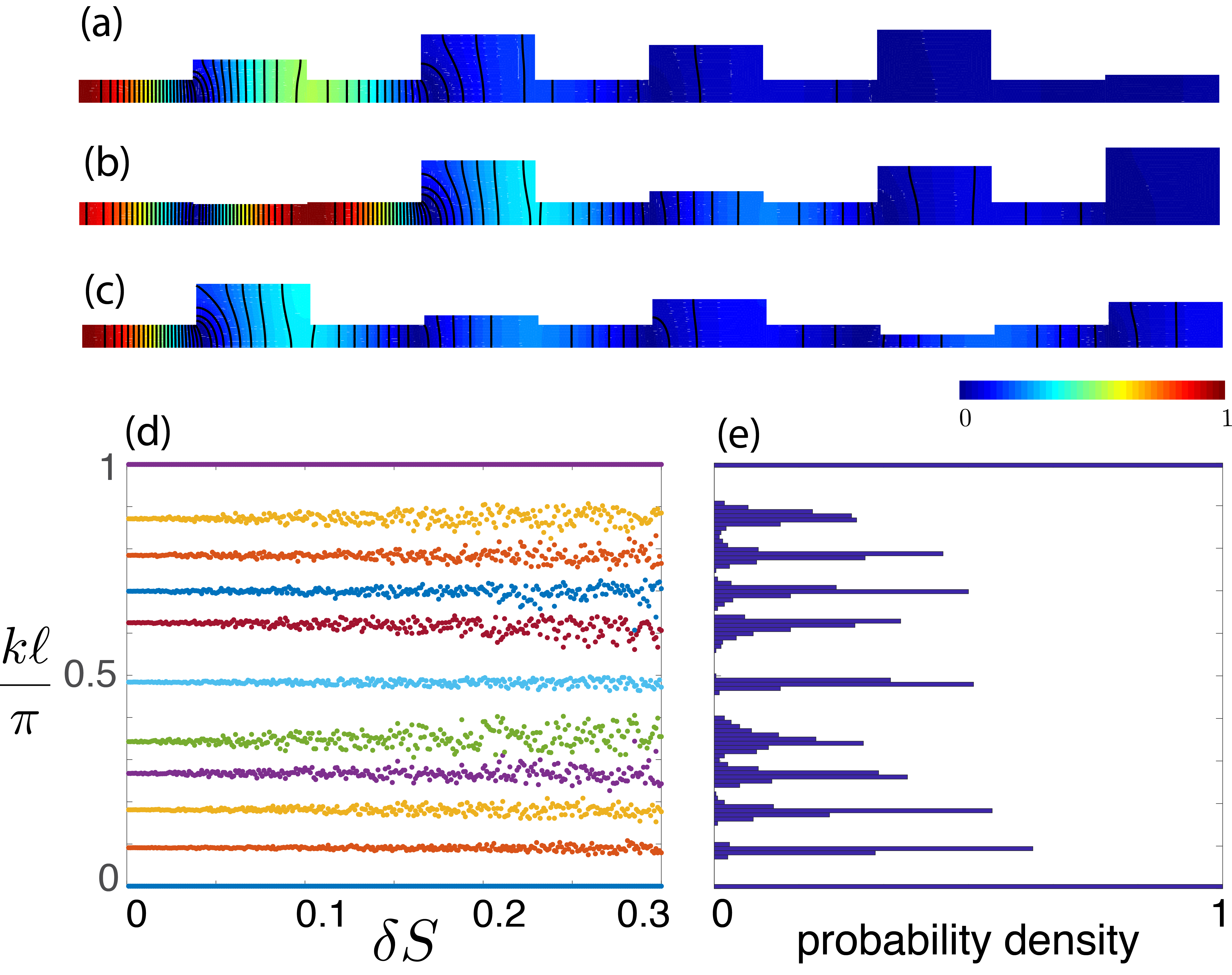}
\caption{\label{fig5} (a) The edge mode in the periodic case (without disorder). (b-d) The edge modes for three different disorder realizations of strength $\delta S=0.3$ as obtained numerically.
(e) The eigenfrequency spectrum of a disorder realization as a function of the disorder strength $\delta S$. (f) The probability density of the eigenvalues for 
a set of $10^3$ disorder realizations. } 
\end{figure}

Now, the set of equations~\eqref{DisorderSSH_nonNorm1} and \eqref{DisorderSSH_nonNorm2} presents a technical difficulty, as it does not define a hermitian eigenvalue problem
due to asymmetric coupling coefficients.
 To remedy this, a solution is to use normalized pressure values 
 instead of physical ones. For this, we define 
\begin{subequations} \label{NormPressure} \begin{eqnarray}
\tilde A_n &=& \sqrt{S_{2n-1}+S_{2n}} A_n, \\
\tilde B_n &=& \sqrt{S_{2n}+S_{2n+1}} B_n . 
\end{eqnarray} \end{subequations}
The set of equations~\eqref{DisorderSSH_nonNorm1} and \eqref{DisorderSSH_nonNorm2} can now be written as a
 disordered SSH model with symmetric coupling $s_j$: 
\begin{subequations} \label{DisorderSSH} \begin{eqnarray}
E(k) \tilde A_n &=& t_{n-1} \tilde B_{n-1} + s_{n} \tilde B_n , \\
E(k) \tilde B_n &=& s_n \tilde A_{n} + t_n \tilde A_{n+1} , 
\end{eqnarray} \end{subequations}
with 
\begin{subequations} \label{DisorderSSH_Hoppings} \begin{eqnarray}
s_n &=& \frac{S_{2n}}{\sqrt{(S_{2n-1}+S_{2n})(S_{2n}+S_{2n+1})}} , \\
t_n &=& \frac{S_{2n+1}}{\sqrt{(S_{2n+1}+S_{2n+2})(S_{2n}+S_{2n+1})}} . 
\end{eqnarray} \end{subequations}
We now compute the spectrum of a finite disordered waveguide with $2N+1$ cross section changes and closed boundary conditions. We take a fixed value of the cross section $S_{2n-1} = S_A$, and the values of $S_{2n}$ are uniformly random in the interval $[S_B-\delta S,S_B+\delta S]$. The results are shown in Fig.~\ref{fig5}. 
For 3 different realisations of strong disorder, from finite element solutions 
of the Helmholtz equation, we see that the edge mode is preserved (Fig.~\ref{fig5}(a-c)). 
The evolution  of eigenvalues as the strength of disorder $\delta S$ is increased
is displayed in Fig.~\ref{fig5}(d).  We see that the edge mode eigenfrequency (always close to $kl=\pi / 2$) is much more robust compared to the neibourghing  bulk modes. 
We note here that the for the discrete equations \eqref{DisorderSSH}, the edge mode energy is exactly $E=0$ in a chiral disorder, hence $k\ell = \pi/2$. The small frequency deviations seen in the 2D model (Fig.~\ref{fig5}(d,e)) originate from the 1D monomode approximation which is not fully valid for some values of the cross ratio; see for example the curved contour pressure lines in Fig.~\ref{fig5}(c). A particularity of the setup is that modes at $k l=0$ and $k l=\pi$ are exact solutions of the 2D wave equation, making 
these eigenfrequencies insensitive to the chosen disorder. 
To get a clearer view of the effect of disorder the probability distribution of the 
eigenfrequencies are shown in Fig.~\ref{fig5}(e). 

Interestingly, for an odd number of cross section changes, the edge mode has a closed form expression similar to equation~\eqref{Periodic_EdgeMode_CloseOdd}~\cite{Coutant20}. Indeed, equations \eqref{DisorderSSH} for $E=0$ are solved by 
\begin{equation} \label{Disorder_EdgeMode_CloseOdd}
\bmat \tilde A_n \\ \tilde B_n \emat = \alpha_0 \bmat 0 \\ 1 \emat \prod_{j=1}^n \left(-t_j/s_j \right) . 
\end{equation}
Notice that this is simply equation~\eqref{Periodic_EdgeMode_CloseOdd} with varying hopping coefficients. This expression has various interesting consequences. In the limit of semi-infinite system ($N \to \infty$), one can define the localization length of the zero-energy mode as 
\begin{equation}
L^{-1} = - \lim_{n \to \infty} \frac1n \ln \left( \max \left( |\tilde A_n| ; |\tilde B_n| \right) \right), 
\end{equation}
which from equation \eqref{Disorder_EdgeMode_CloseOdd} gives 
\begin{equation} \label{L_rw}
L^{-1} = \lim_{n\to \infty} \frac1n \sum_{j=1}^n \ln \left|\frac{t_j}{s_j} \right| . 
\end{equation}
At this level, the random variable $\ln \left|t_j/s_j \right|$ can be seen as a (biased) random walk~\cite{Inui94,Mondragon14}. Using the ergodic theorem, the spatial average \eqref{L_rw} (i.e. time average in the analogue random walk) can be replaced by an average over disorder realisations. Therefore, the localization length is given by 
\begin{equation}
L^{-1} = \langle \ln \left| t / s \right| \rangle = \langle \ln \left| t \right| \rangle - \langle \ln \left| s \right| \rangle, 
\end{equation}
where $\langle . \rangle$ means average over disorder realisations. Notice that $L$ is finite whenever $\langle \ln \left| s \right| \rangle < \langle \ln \left| t \right| \rangle$. When $\langle \ln \left| s \right| \rangle = \langle \ln \left| t \right| \rangle$ we have $L=\infty$ but this does not mean the zero energy mode is extended. In fact, the zero-energy level for $\langle \ln \left| s \right| \rangle = \langle \ln \left| t \right| \rangle$ is localized but decreases as $O(e^{-\lambda \sqrt{n}})$~\cite{Inui94}. 

\section{Experiments}
\label{Exp_Sec}
\begin{figure}[t!]
\includegraphics[width=\columnwidth]{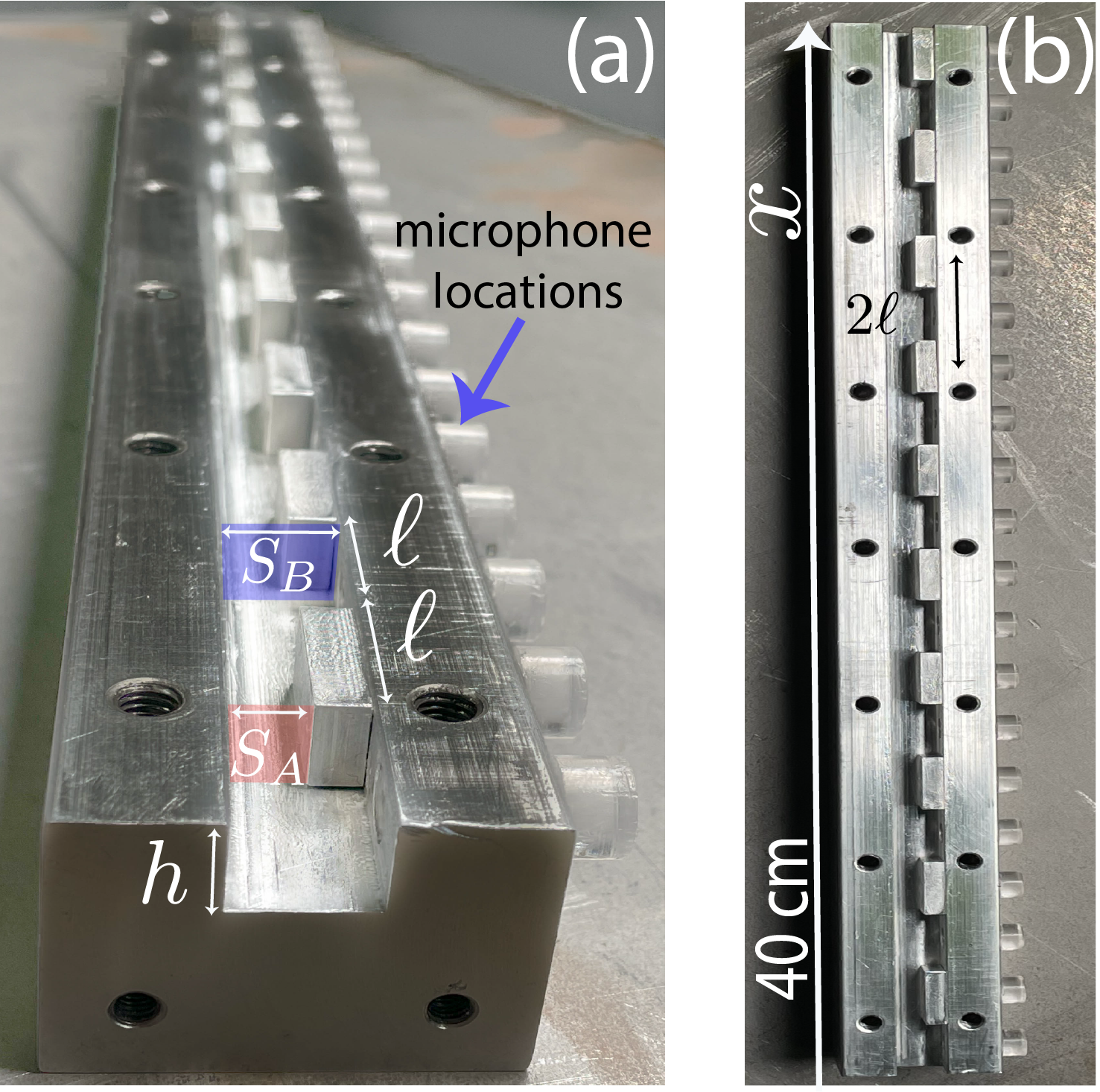}
\caption{Experimental setup. (a) Side view of the waveguide  where the two different crossections
are indicated. Cross section changes appear every $\ell$.   In (b) we show the whole waveguide of total  length $40$~cm
which is composed of $N=10$ unit cells. }
\label{Fig5_exp}
\end{figure}

Our experimental setup is based on an acoustic waveguide of rectangular cross section with a constant height of $ 1$~cm. 
The 1D SSH lattice  is realized by changing the cross section of the main waveguide every $\ell=2$~cm.
This is done by adding aluminium blocks as shown in Fig.~\ref{Fig5_exp} where an example of a specific choice of periodic configuration with
2 different cross sections $S_A$ and $S_B$ is illustrated.  In the experiments, the waveguide is closed on the top by plexiglass plate.
Measurements of the acoustic pressure are done using microphones at the different positions indicated in Fig.~\ref{Fig5_exp}.
Here, we choose to work with closed boundary conditions which is achieved by blocking the two ends of the waveguide
using plexiglass plates.

The total length of the waveguide is $40$~cm. Since $\ell=2$~cm and due to the fact that we use closed boundary conditions this corresponds to a SSH lattice model of 21 lattice sites (see discussion in Section III and Fig.~\ref{Fig5_exp}). To verify the appearance of the edge mode in the acoustic system, we performed experiments using two different cross sections $S_A=0.5\times 1$~cm$^2$ and $S_B=0.85\times 1$~cm$^2$ corresponding to $t=0.63=1-s$. Note that according to our analytical model the frequency of the edge mode is simply given by $k\ell=\pi/2$ which for our setup corresponds to $f_0=4.2875$~kHz assuming a speed of sound in air of $c=343$~m.s$^{-1}$.

\begin{figure}[htp]
\includegraphics[width=8.5cm]{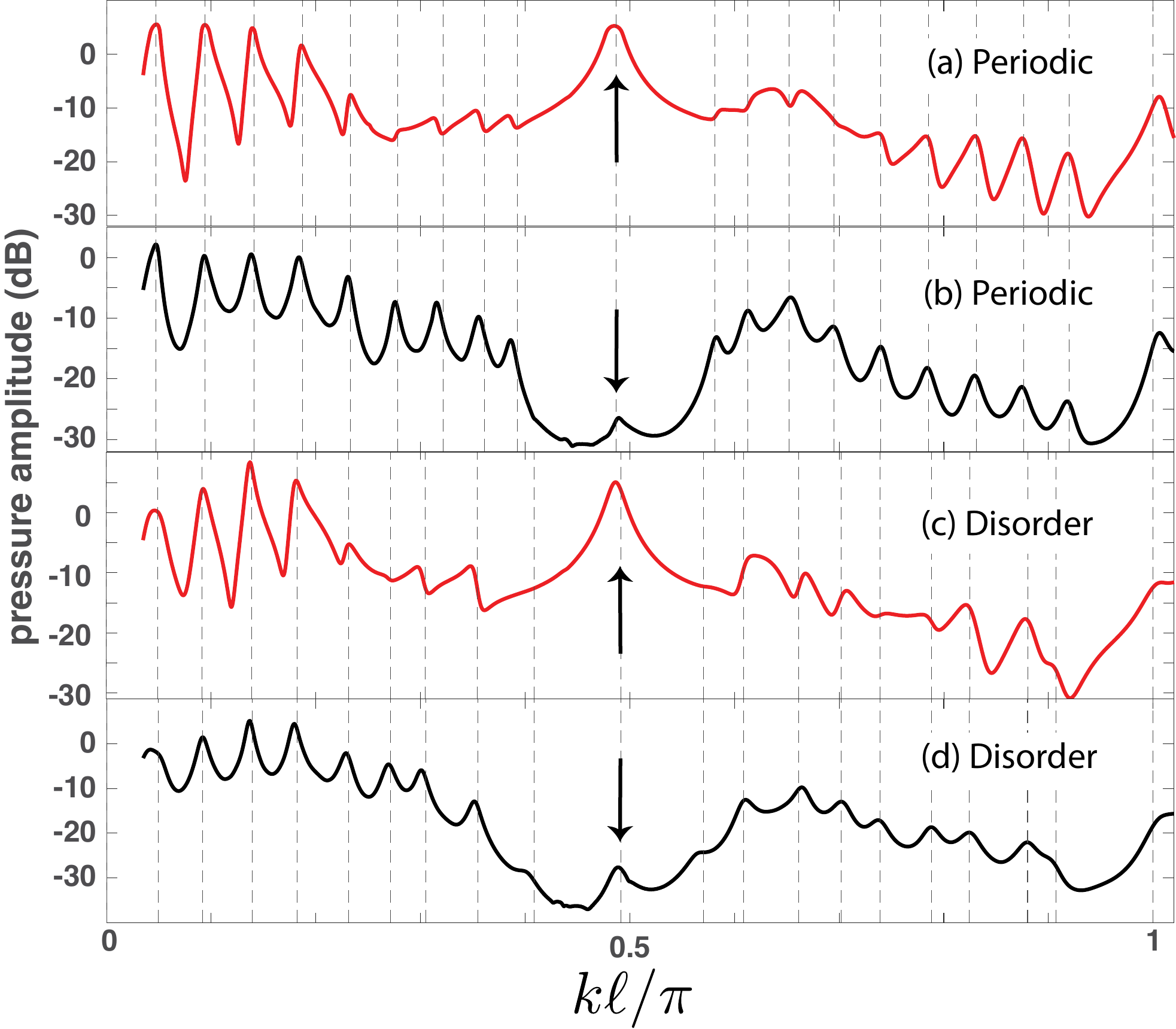}
\caption{(a) Pressure amplitude as a function of frequency for a periodic SSH configuration with $t=0.63$ measured at $4$~cm from the source.   (b) Same as panel (a) but the pressure amplitude is obtained using the mean value of the measured pressure at all the junctions. The dashed vertical lines indicate the frequency of the modes as obtained numerically. (c)[(d)] Same as panel (a)[(b)] for a disordered configuration with a mean value $\langle t\rangle=0.63$. }
\label{Fig7a}
\end{figure}

We use a source positioned at the end of the waveguide where the edge mode is localized [top of Fig.~\ref{Fig5_exp} (b)] with a sweep-sine signal. The spectrum as measured at $4$~cm from the source (corresponding to the $19$~th lattice site) is shown in Fig.~\ref{Fig7a}(a). The edge mode
is clearly observed as indicated by the large resonance peak around $k\ell=\pi/2$. The dashed vertical line at this frequency denotes the corresponding numerical result. For completeness, in Fig.~\ref{Fig7a}(b) we show the spectrum after averaging over the different positions, where the peaks of the 21 modes are exposed, and the appearance band gap centered  around  $k\ell=\pi/2$ is evident. Note that, as expected the edge mode lies in the center of this gap (indicated by the arrow).

\begin{figure}[htp]
\includegraphics[width=8.5cm]{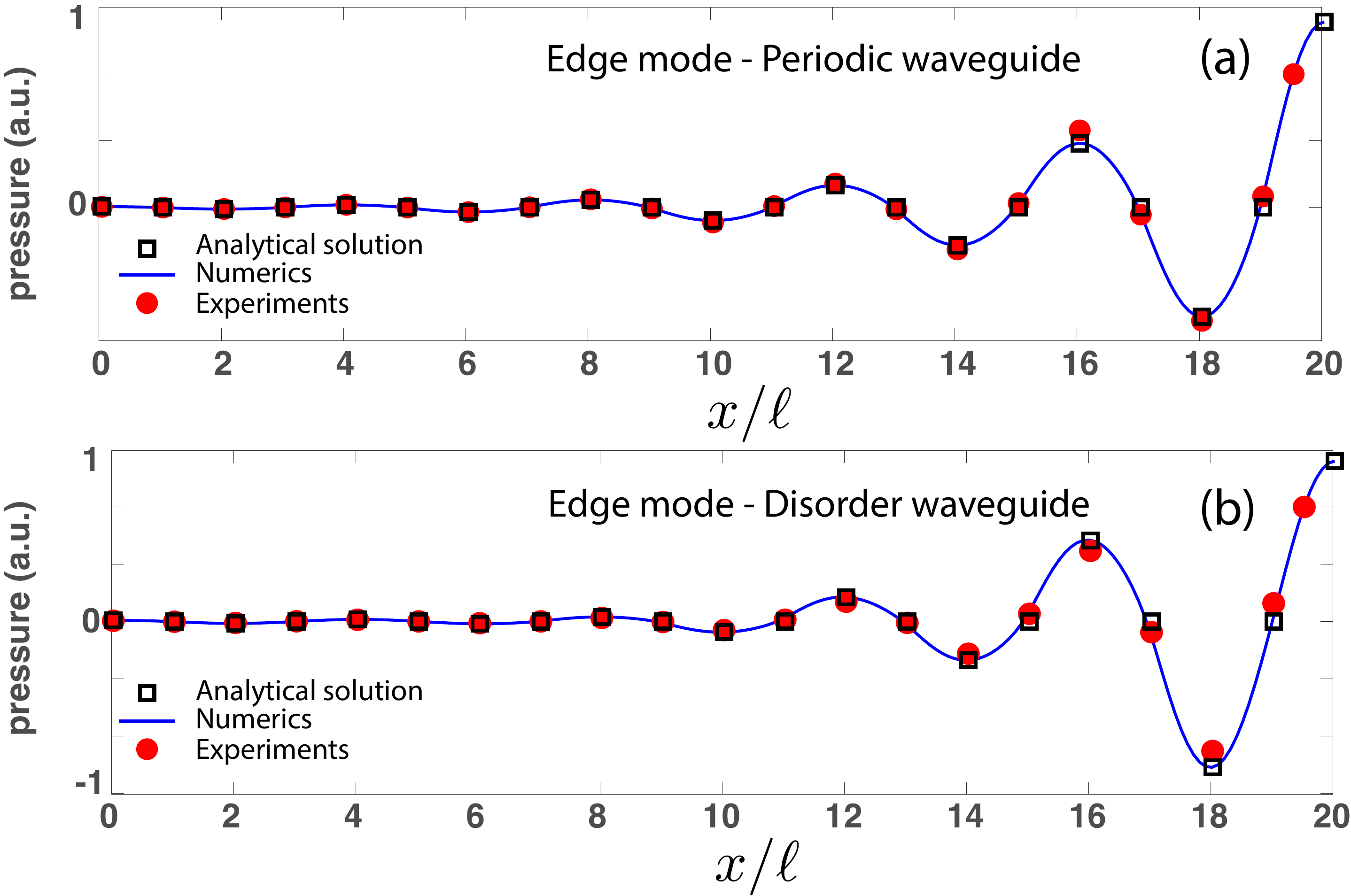}
\caption{(a) The edge mode profile corresponding to the experimental setup as obtained by the analytical solution (squares), numerical simulations (solid line) and experiments (circles). (b) The same as (a) but for the case of a disorder realization.}
\label{Fig7b}
\end{figure}

To further confirm the appearance of the SSH edge mode we also plot the experimentally obtained
profile at the frequency of the peak. The result is shown in Fig.~\ref{Fig7b}(a) by the red circles and it is compared both with the numerical solution (solid line) and the discrete exact analytical solution of equation~(\ref{Periodic_EdgeMode_CloseOdd}) [black squares]. The characteristic profile of the edge mode with vanishing pressure at one sub-lattice (A in our case) is recovered. It is quite remarkable that the profile of the acoustic mode from experiments fits perfectly with the analytical SSH discrete solution \textit{without} any fitting parameter.

Next, we experimentally study the effect of disorder by randomly changing the cross sections $S_B$. In order to compare with the corresponding periodic case of Figs.~\ref{Fig7a}(a) and (b), we choose the cross-sections to be $S_{2n}=S_B(1+\delta S)$ where $\delta S\in [-0.35,0.35]$ is a random number.
A typical example of the spectrum of a disorder realization is shown in Fig.~\ref{Fig7a} (c). It is directly seen that the peak at the center of the gap is robust to the disorder. 
We have measured a frequency shift between the two edge modes of the ordered and disordered lattice to be less than $0.2\%$. On the other hand, the rest of the modes as indicated by the vertical dashed lines are shifted due to disorder. Furthermore, to complete the experimental analysis of the edge mode in the presence of disorder we plot the corresponding profile in Fig.~\ref{Fig7b}(b) confirming the localization of this mode. 
Even in the presence of disorder the exact analytical profile of the mode as given by equation~(\ref{Disorder_EdgeMode_CloseOdd}) captures perfectly the experimental results.

\section{Concluding remarks}

The approach followed here is to begin by finding a discrete modelling of the continuous two-dimensional acoustic wave equation.
It appears that our waveguide setup can be mapped to a dimer discrete model with chiral symmetry (the SSH model), the band gap of 
which is closed at the edge of the Brillouin zone for the trivial case of a uniform waveguide (Fig. 2(b)). 
This is different from the cases of previous acoustic waveguide analogues to SSH model where 
either the TBA with fitted coupling was chosen or fine tuning of the length of the waveguide segments led to band inversion through an accidental degeneracy. 
Our acoustic Su-Schrieffer-Heeger lattice has been validated by comparison with direct numerical computations and experimental measurements.
As a consequence of the underlying 
chiral symmetry in our case, a topological edge mode is obtained in every band gap of the continuum model, 
without the need to calculate the corresponding Zak phases of the bands, since the topological characterization is directly inherited from the SSH model. 
We believe that the simplicity and the versatility of our acoustic SSH platform
offers new opportunities to inspect topological effects for acoustic waves.

\section*{Aknowledgements}

This project has received funding from the European Union's Horizon 2020 research and innovation programme under the Marie Sklodowska-Curie grant agreement No 843152. 
G.T. acknowledges financial support from the CS.MICRO project funded under the program Etoiles Montantes of the Region Pays de la Loire.
V.A. acknowledges financial support from the NoHENA project funded under the program Etoiles Montantes of the Region Pays de la Loire.

\bibliographystyle{apsrev4-1}
\bibliography{Bibli}

\begin{thebibliography}{29}%
\makeatletter
\providecommand \@ifxundefined [1]{%
 \@ifx{#1\undefined}
}%
\providecommand \@ifnum [1]{%
 \ifnum #1\expandafter \@firstoftwo
 \else \expandafter \@secondoftwo
 \fi
}%
\providecommand \@ifx [1]{%
 \ifx #1\expandafter \@firstoftwo
 \else \expandafter \@secondoftwo
 \fi
}%
\providecommand \natexlab [1]{#1}%
\providecommand \enquote  [1]{``#1''}%
\providecommand \bibnamefont  [1]{#1}%
\providecommand \bibfnamefont [1]{#1}%
\providecommand \citenamefont [1]{#1}%
\providecommand \href@noop [0]{\@secondoftwo}%
\providecommand \href [0]{\begingroup \@sanitize@url \@href}%
\providecommand \@href[1]{\@@startlink{#1}\@@href}%
\providecommand \@@href[1]{\endgroup#1\@@endlink}%
\providecommand \@sanitize@url [0]{\catcode `\\12\catcode `\$12\catcode
  `\&12\catcode `\#12\catcode `\^12\catcode `\_12\catcode `\%12\relax}%
\providecommand \@@startlink[1]{}%
\providecommand \@@endlink[0]{}%
\providecommand \url  [0]{\begingroup\@sanitize@url \@url }%
\providecommand \@url [1]{\endgroup\@href {#1}{\urlprefix }}%
\providecommand \urlprefix  [0]{URL }%
\providecommand \Eprint [0]{\href }%
\providecommand \doibase [0]{http://dx.doi.org/}%
\providecommand \selectlanguage [0]{\@gobble}%
\providecommand \bibinfo  [0]{\@secondoftwo}%
\providecommand \bibfield  [0]{\@secondoftwo}%
\providecommand \translation [1]{[#1]}%
\providecommand \BibitemOpen [0]{}%
\providecommand \bibitemStop [0]{}%
\providecommand \bibitemNoStop [0]{.\EOS\space}%
\providecommand \EOS [0]{\spacefactor3000\relax}%
\providecommand \BibitemShut  [1]{\csname bibitem#1\endcsname}%
\let\auto@bib@innerbib\@empty
\bibitem [{\citenamefont {Thouless}\ \emph {et~al.}(1982)\citenamefont
  {Thouless}, \citenamefont {Kohmoto}, \citenamefont {Nightingale},\ and\
  \citenamefont {den Nijs}}]{Thouless82}%
  \BibitemOpen
  \bibfield  {author} {\bibinfo {author} {\bibfnamefont {D.~J.}\ \bibnamefont
  {Thouless}}, \bibinfo {author} {\bibfnamefont {M.}~\bibnamefont {Kohmoto}},
  \bibinfo {author} {\bibfnamefont {M.~P.}\ \bibnamefont {Nightingale}}, \ and\
  \bibinfo {author} {\bibfnamefont {M.}~\bibnamefont {den Nijs}},\ }\href@noop
  {} {\bibfield  {journal} {\bibinfo  {journal} {Phys. Rev. Lett.}\ }\textbf
  {\bibinfo {volume} {49}},\ \bibinfo {pages} {405} (\bibinfo {year}
  {1982})}\BibitemShut {NoStop}%
\bibitem [{\citenamefont {Ozawa}\ \emph {et~al.}(2019)\citenamefont {Ozawa},
  \citenamefont {Price}, \citenamefont {Amo}, \citenamefont {Goldman},
  \citenamefont {Hafezi}, \citenamefont {Lu}, \citenamefont {Rechtsman},
  \citenamefont {Schuster}, \citenamefont {Simon}, \citenamefont {Zilberberg}
  \emph {et~al.}}]{Ozawa19}%
  \BibitemOpen
  \bibfield  {author} {\bibinfo {author} {\bibfnamefont {T.}~\bibnamefont
  {Ozawa}}, \bibinfo {author} {\bibfnamefont {H.~M.}\ \bibnamefont {Price}},
  \bibinfo {author} {\bibfnamefont {A.}~\bibnamefont {Amo}}, \bibinfo {author}
  {\bibfnamefont {N.}~\bibnamefont {Goldman}}, \bibinfo {author} {\bibfnamefont
  {M.}~\bibnamefont {Hafezi}}, \bibinfo {author} {\bibfnamefont
  {L.}~\bibnamefont {Lu}}, \bibinfo {author} {\bibfnamefont {M.~C.}\
  \bibnamefont {Rechtsman}}, \bibinfo {author} {\bibfnamefont {D.}~\bibnamefont
  {Schuster}}, \bibinfo {author} {\bibfnamefont {J.}~\bibnamefont {Simon}},
  \bibinfo {author} {\bibfnamefont {O.}~\bibnamefont {Zilberberg}},  \emph
  {et~al.},\ }\href@noop {} {\bibfield  {journal} {\bibinfo  {journal} {Reviews
  of Modern Physics}\ }\textbf {\bibinfo {volume} {91}},\ \bibinfo {pages}
  {015006} (\bibinfo {year} {2019})}\BibitemShut {NoStop}%
\bibitem [{\citenamefont {Huber}(2016)}]{Huber16}%
  \BibitemOpen
  \bibfield  {author} {\bibinfo {author} {\bibfnamefont {S.~D.}\ \bibnamefont
  {Huber}},\ }\href@noop {} {\bibfield  {journal} {\bibinfo  {journal} {Nature
  Physics}\ }\textbf {\bibinfo {volume} {12}},\ \bibinfo {pages} {621}
  (\bibinfo {year} {2016})}\BibitemShut {NoStop}%
\bibitem [{\citenamefont {Zhang}\ \emph {et~al.}(2018)\citenamefont {Zhang},
  \citenamefont {Xiao}, \citenamefont {Cheng}, \citenamefont {Lu},\ and\
  \citenamefont {Christensen}}]{Zhang18}%
  \BibitemOpen
  \bibfield  {author} {\bibinfo {author} {\bibfnamefont {X.}~\bibnamefont
  {Zhang}}, \bibinfo {author} {\bibfnamefont {M.}~\bibnamefont {Xiao}},
  \bibinfo {author} {\bibfnamefont {Y.}~\bibnamefont {Cheng}}, \bibinfo
  {author} {\bibfnamefont {M.-H.}\ \bibnamefont {Lu}}, \ and\ \bibinfo {author}
  {\bibfnamefont {J.}~\bibnamefont {Christensen}},\ }\href@noop {} {\bibfield
  {journal} {\bibinfo  {journal} {Communications Physics}\ }\textbf {\bibinfo
  {volume} {1}},\ \bibinfo {pages} {1} (\bibinfo {year} {2018})}\BibitemShut
  {NoStop}%
\bibitem [{\citenamefont {Ma}\ \emph {et~al.}(2019)\citenamefont {Ma},
  \citenamefont {Xiao},\ and\ \citenamefont {Chan}}]{Ma19}%
  \BibitemOpen
  \bibfield  {author} {\bibinfo {author} {\bibfnamefont {G.}~\bibnamefont
  {Ma}}, \bibinfo {author} {\bibfnamefont {M.}~\bibnamefont {Xiao}}, \ and\
  \bibinfo {author} {\bibfnamefont {C.~T.}\ \bibnamefont {Chan}},\ }\href@noop
  {} {\bibfield  {journal} {\bibinfo  {journal} {Nature Reviews Physics}\
  }\textbf {\bibinfo {volume} {1}},\ \bibinfo {pages} {281} (\bibinfo {year}
  {2019})}\BibitemShut {NoStop}%
\bibitem [{\citenamefont {Su}\ \emph {et~al.}(1979)\citenamefont {Su},
  \citenamefont {Schrieffer},\ and\ \citenamefont {Heeger}}]{Su79}%
  \BibitemOpen
  \bibfield  {author} {\bibinfo {author} {\bibfnamefont {W.}~\bibnamefont
  {Su}}, \bibinfo {author} {\bibfnamefont {J.}~\bibnamefont {Schrieffer}}, \
  and\ \bibinfo {author} {\bibfnamefont {A.~J.}\ \bibnamefont {Heeger}},\
  }\href@noop {} {\bibfield  {journal} {\bibinfo  {journal} {Phys. Rev. Lett.}\
  }\textbf {\bibinfo {volume} {42}},\ \bibinfo {pages} {1698} (\bibinfo {year}
  {1979})}\BibitemShut {NoStop}%
\bibitem [{\citenamefont {Asb{\'o}th}\ \emph {et~al.}(2016)\citenamefont
  {Asb{\'o}th}, \citenamefont {Oroszl{\'a}ny},\ and\ \citenamefont
  {P{\'a}lyi}}]{Asboth16}%
  \BibitemOpen
  \bibfield  {author} {\bibinfo {author} {\bibfnamefont {J.~K.}\ \bibnamefont
  {Asb{\'o}th}}, \bibinfo {author} {\bibfnamefont {L.}~\bibnamefont
  {Oroszl{\'a}ny}}, \ and\ \bibinfo {author} {\bibfnamefont {A.}~\bibnamefont
  {P{\'a}lyi}},\ }\href@noop {} {\bibfield  {journal} {\bibinfo  {journal}
  {Lecture notes in physics}\ }\textbf {\bibinfo {volume} {919}},\ \bibinfo
  {pages} {87} (\bibinfo {year} {2016})}\BibitemShut {NoStop}%
\bibitem [{\citenamefont {Yang}\ and\ \citenamefont
  {Zhang}(2016)}]{yang2016acoustic}%
  \BibitemOpen
  \bibfield  {author} {\bibinfo {author} {\bibfnamefont {Z.}~\bibnamefont
  {Yang}}\ and\ \bibinfo {author} {\bibfnamefont {B.}~\bibnamefont {Zhang}},\
  }\href@noop {} {\bibfield  {journal} {\bibinfo  {journal} {Physical review
  letters}\ }\textbf {\bibinfo {volume} {117}},\ \bibinfo {pages} {224301}
  (\bibinfo {year} {2016})}\BibitemShut {NoStop}%
\bibitem [{\citenamefont {Li}\ \emph {et~al.}(2018)\citenamefont {Li},
  \citenamefont {Meng}, \citenamefont {Wu}, \citenamefont {Yan}, \citenamefont
  {Huang}, \citenamefont {Wang},\ and\ \citenamefont {Wen}}]{Li18}%
  \BibitemOpen
  \bibfield  {author} {\bibinfo {author} {\bibfnamefont {X.}~\bibnamefont
  {Li}}, \bibinfo {author} {\bibfnamefont {Y.}~\bibnamefont {Meng}}, \bibinfo
  {author} {\bibfnamefont {X.}~\bibnamefont {Wu}}, \bibinfo {author}
  {\bibfnamefont {S.}~\bibnamefont {Yan}}, \bibinfo {author} {\bibfnamefont
  {Y.}~\bibnamefont {Huang}}, \bibinfo {author} {\bibfnamefont
  {S.}~\bibnamefont {Wang}}, \ and\ \bibinfo {author} {\bibfnamefont
  {W.}~\bibnamefont {Wen}},\ }\href@noop {} {\bibfield  {journal} {\bibinfo
  {journal} {Applied Physics Letters}\ }\textbf {\bibinfo {volume} {113}},\
  \bibinfo {pages} {203501} (\bibinfo {year} {2018})},\ \Eprint
  {http://arxiv.org/abs/1808.06949} {arXiv:1808.06949 [physics.class-ph]}
  \BibitemShut {NoStop}%
\bibitem [{\citenamefont {Esmann}\ \emph {et~al.}(2018)\citenamefont {Esmann},
  \citenamefont {Lamberti}, \citenamefont {Lema{\^\i}tre},\ and\ \citenamefont
  {Lanzillotti-Kimura}}]{esmann2018topological}%
  \BibitemOpen
  \bibfield  {author} {\bibinfo {author} {\bibfnamefont {M.}~\bibnamefont
  {Esmann}}, \bibinfo {author} {\bibfnamefont {F.}~\bibnamefont {Lamberti}},
  \bibinfo {author} {\bibfnamefont {A.}~\bibnamefont {Lema{\^\i}tre}}, \ and\
  \bibinfo {author} {\bibfnamefont {N.}~\bibnamefont {Lanzillotti-Kimura}},\
  }\href@noop {} {\bibfield  {journal} {\bibinfo  {journal} {Physical Review
  B}\ }\textbf {\bibinfo {volume} {98}},\ \bibinfo {pages} {161109} (\bibinfo
  {year} {2018})}\BibitemShut {NoStop}%
\bibitem [{\citenamefont {Shen}\ \emph {et~al.}(2020)\citenamefont {Shen},
  \citenamefont {Zeng}, \citenamefont {Geng}, \citenamefont {Zhao},
  \citenamefont {Peng},\ and\ \citenamefont {Zhu}}]{shen2020acoustic}%
  \BibitemOpen
  \bibfield  {author} {\bibinfo {author} {\bibfnamefont {Y.-X.}\ \bibnamefont
  {Shen}}, \bibinfo {author} {\bibfnamefont {L.-S.}\ \bibnamefont {Zeng}},
  \bibinfo {author} {\bibfnamefont {Z.-G.}\ \bibnamefont {Geng}}, \bibinfo
  {author} {\bibfnamefont {D.-G.}\ \bibnamefont {Zhao}}, \bibinfo {author}
  {\bibfnamefont {Y.-G.}\ \bibnamefont {Peng}}, \ and\ \bibinfo {author}
  {\bibfnamefont {X.-F.}\ \bibnamefont {Zhu}},\ }\href@noop {} {\bibfield
  {journal} {\bibinfo  {journal} {Physical Review Applied}\ }\textbf {\bibinfo
  {volume} {14}},\ \bibinfo {pages} {014043} (\bibinfo {year}
  {2020})}\BibitemShut {NoStop}%
\bibitem [{\citenamefont {Yan}\ \emph {et~al.}(2020)\citenamefont {Yan},
  \citenamefont {Huang}, \citenamefont {Luo}, \citenamefont {Lu}, \citenamefont
  {Deng},\ and\ \citenamefont {Liu}}]{yan2020acoustic}%
  \BibitemOpen
  \bibfield  {author} {\bibinfo {author} {\bibfnamefont {M.}~\bibnamefont
  {Yan}}, \bibinfo {author} {\bibfnamefont {X.}~\bibnamefont {Huang}}, \bibinfo
  {author} {\bibfnamefont {L.}~\bibnamefont {Luo}}, \bibinfo {author}
  {\bibfnamefont {J.}~\bibnamefont {Lu}}, \bibinfo {author} {\bibfnamefont
  {W.}~\bibnamefont {Deng}}, \ and\ \bibinfo {author} {\bibfnamefont
  {Z.}~\bibnamefont {Liu}},\ }\href@noop {} {\bibfield  {journal} {\bibinfo
  {journal} {Physical Review B}\ }\textbf {\bibinfo {volume} {102}},\ \bibinfo
  {pages} {180102} (\bibinfo {year} {2020})}\BibitemShut {NoStop}%
\bibitem [{\citenamefont {Chen}\ \emph {et~al.}(2020)\citenamefont {Chen},
  \citenamefont {Wang}, \citenamefont {Zhang},\ and\ \citenamefont
  {Ma}}]{chen2020chiral}%
  \BibitemOpen
  \bibfield  {author} {\bibinfo {author} {\bibfnamefont {Z.-G.}\ \bibnamefont
  {Chen}}, \bibinfo {author} {\bibfnamefont {L.}~\bibnamefont {Wang}}, \bibinfo
  {author} {\bibfnamefont {G.}~\bibnamefont {Zhang}}, \ and\ \bibinfo {author}
  {\bibfnamefont {G.}~\bibnamefont {Ma}},\ }\href@noop {} {\bibfield  {journal}
  {\bibinfo  {journal} {Physical Review Applied}\ }\textbf {\bibinfo {volume}
  {14}},\ \bibinfo {pages} {024023} (\bibinfo {year} {2020})}\BibitemShut
  {NoStop}%
\bibitem [{\citenamefont {Xiao}\ \emph {et~al.}(2015)\citenamefont {Xiao},
  \citenamefont {Ma}, \citenamefont {Yang}, \citenamefont {Sheng},
  \citenamefont {Zhang},\ and\ \citenamefont {Chan}}]{Xiao15}%
  \BibitemOpen
  \bibfield  {author} {\bibinfo {author} {\bibfnamefont {M.}~\bibnamefont
  {Xiao}}, \bibinfo {author} {\bibfnamefont {G.}~\bibnamefont {Ma}}, \bibinfo
  {author} {\bibfnamefont {Z.}~\bibnamefont {Yang}}, \bibinfo {author}
  {\bibfnamefont {P.}~\bibnamefont {Sheng}}, \bibinfo {author} {\bibfnamefont
  {Z.}~\bibnamefont {Zhang}}, \ and\ \bibinfo {author} {\bibfnamefont {C.~T.}\
  \bibnamefont {Chan}},\ }\href@noop {} {\bibfield  {journal} {\bibinfo
  {journal} {Nature Physics}\ }\textbf {\bibinfo {volume} {11}},\ \bibinfo
  {pages} {240} (\bibinfo {year} {2015})}\BibitemShut {NoStop}%
\bibitem [{\citenamefont {Meng}\ \emph {et~al.}(2018)\citenamefont {Meng},
  \citenamefont {Wu}, \citenamefont {Zhang}, \citenamefont {Li}, \citenamefont
  {Hu}, \citenamefont {Ge}, \citenamefont {Huang}, \citenamefont {Xiang},
  \citenamefont {Han}, \citenamefont {Wang} \emph {et~al.}}]{Meng18}%
  \BibitemOpen
  \bibfield  {author} {\bibinfo {author} {\bibfnamefont {Y.}~\bibnamefont
  {Meng}}, \bibinfo {author} {\bibfnamefont {X.}~\bibnamefont {Wu}}, \bibinfo
  {author} {\bibfnamefont {R.-Y.}\ \bibnamefont {Zhang}}, \bibinfo {author}
  {\bibfnamefont {X.}~\bibnamefont {Li}}, \bibinfo {author} {\bibfnamefont
  {P.}~\bibnamefont {Hu}}, \bibinfo {author} {\bibfnamefont {L.}~\bibnamefont
  {Ge}}, \bibinfo {author} {\bibfnamefont {Y.}~\bibnamefont {Huang}}, \bibinfo
  {author} {\bibfnamefont {H.}~\bibnamefont {Xiang}}, \bibinfo {author}
  {\bibfnamefont {D.}~\bibnamefont {Han}}, \bibinfo {author} {\bibfnamefont
  {S.}~\bibnamefont {Wang}},  \emph {et~al.},\ }\href@noop {} {\bibfield
  {journal} {\bibinfo  {journal} {New Journal of Physics}\ }\textbf {\bibinfo
  {volume} {20}},\ \bibinfo {pages} {073032} (\bibinfo {year} {2018})},\
  \Eprint {http://arxiv.org/abs/1804.10754} {arXiv:1804.10754 [physics.app-ph]}
  \BibitemShut {NoStop}%
\bibitem [{\citenamefont {Yin}\ \emph {et~al.}(2018)\citenamefont {Yin},
  \citenamefont {Ruzzene}, \citenamefont {Wen}, \citenamefont {Yu},
  \citenamefont {Cai},\ and\ \citenamefont {Yue}}]{yin2018band}%
  \BibitemOpen
  \bibfield  {author} {\bibinfo {author} {\bibfnamefont {J.}~\bibnamefont
  {Yin}}, \bibinfo {author} {\bibfnamefont {M.}~\bibnamefont {Ruzzene}},
  \bibinfo {author} {\bibfnamefont {J.}~\bibnamefont {Wen}}, \bibinfo {author}
  {\bibfnamefont {D.}~\bibnamefont {Yu}}, \bibinfo {author} {\bibfnamefont
  {L.}~\bibnamefont {Cai}}, \ and\ \bibinfo {author} {\bibfnamefont
  {L.}~\bibnamefont {Yue}},\ }\href@noop {} {\bibfield  {journal} {\bibinfo
  {journal} {Scientific reports}\ }\textbf {\bibinfo {volume} {8}},\ \bibinfo
  {pages} {1} (\bibinfo {year} {2018})}\BibitemShut {NoStop}%
\bibitem [{\citenamefont {Zangeneh-Nejad}\ and\ \citenamefont
  {Fleury}(2019)}]{Zangeneh19}%
  \BibitemOpen
  \bibfield  {author} {\bibinfo {author} {\bibfnamefont {F.}~\bibnamefont
  {Zangeneh-Nejad}}\ and\ \bibinfo {author} {\bibfnamefont {R.}~\bibnamefont
  {Fleury}},\ }\href@noop {} {\bibfield  {journal} {\bibinfo  {journal} {Phys.
  Rev. Lett.}\ }\textbf {\bibinfo {volume} {122}},\ \bibinfo {pages} {014301}
  (\bibinfo {year} {2019})},\ \Eprint {http://arxiv.org/abs/1809.05389}
  {arXiv:1809.05389 [physics.app-ph]} \BibitemShut {NoStop}%
\bibitem [{\citenamefont {Zangeneh-Nejad}\ and\ \citenamefont
  {Fleury}(2020)}]{Zangeneh20}%
  \BibitemOpen
  \bibfield  {author} {\bibinfo {author} {\bibfnamefont {F.}~\bibnamefont
  {Zangeneh-Nejad}}\ and\ \bibinfo {author} {\bibfnamefont {R.}~\bibnamefont
  {Fleury}},\ }\href@noop {} {\bibfield  {journal} {\bibinfo  {journal}
  {Advanced Materials}\ }\textbf {\bibinfo {volume} {32}},\ \bibinfo {pages}
  {2001034} (\bibinfo {year} {2020})}\BibitemShut {NoStop}%
\bibitem [{\citenamefont {Dalmont}\ and\ \citenamefont
  {Kergomard}(1994)}]{Dalmont94}%
  \BibitemOpen
  \bibfield  {author} {\bibinfo {author} {\bibfnamefont {J.-P.}\ \bibnamefont
  {Dalmont}}\ and\ \bibinfo {author} {\bibfnamefont {J.}~\bibnamefont
  {Kergomard}},\ }\href@noop {} {\bibfield  {journal} {\bibinfo  {journal}
  {Acta Acustica}\ }\textbf {\bibinfo {volume} {2}},\ \bibinfo {pages} {421}
  (\bibinfo {year} {1994})}\BibitemShut {NoStop}%
\bibitem [{\citenamefont {Dalmont}\ and\ \citenamefont
  {Vey}(2017)}]{Dalmont17}%
  \BibitemOpen
  \bibfield  {author} {\bibinfo {author} {\bibfnamefont {J.-P.}\ \bibnamefont
  {Dalmont}}\ and\ \bibinfo {author} {\bibfnamefont {G.~L.}\ \bibnamefont
  {Vey}},\ }\href@noop {} {\bibfield  {journal} {\bibinfo  {journal} {Acta
  Acustica united with Acustica}\ }\textbf {\bibinfo {volume} {103}},\ \bibinfo
  {pages} {94} (\bibinfo {year} {2017})}\BibitemShut {NoStop}%
\bibitem [{\citenamefont {Zheng}\ \emph {et~al.}(2019)\citenamefont {Zheng},
  \citenamefont {Achilleos}, \citenamefont {Richoux}, \citenamefont
  {Theocharis},\ and\ \citenamefont {Pagneux}}]{Zheng19}%
  \BibitemOpen
  \bibfield  {author} {\bibinfo {author} {\bibfnamefont {L.-Y.}\ \bibnamefont
  {Zheng}}, \bibinfo {author} {\bibfnamefont {V.}~\bibnamefont {Achilleos}},
  \bibinfo {author} {\bibfnamefont {O.}~\bibnamefont {Richoux}}, \bibinfo
  {author} {\bibfnamefont {G.}~\bibnamefont {Theocharis}}, \ and\ \bibinfo
  {author} {\bibfnamefont {V.}~\bibnamefont {Pagneux}},\ }\href@noop {}
  {\bibfield  {journal} {\bibinfo  {journal} {Physical Review Applied}\
  }\textbf {\bibinfo {volume} {12}},\ \bibinfo {pages} {034014} (\bibinfo
  {year} {2019})}\BibitemShut {NoStop}%
\bibitem [{\citenamefont {Zheng}\ \emph {et~al.}(2020)\citenamefont {Zheng},
  \citenamefont {Achilleos}, \citenamefont {Chen}, \citenamefont {Richoux},
  \citenamefont {Theocharis}, \citenamefont {Wu}, \citenamefont {Mei},
  \citenamefont {Felix}, \citenamefont {Tournat},\ and\ \citenamefont
  {Pagneux}}]{Zheng20}%
  \BibitemOpen
  \bibfield  {author} {\bibinfo {author} {\bibfnamefont {L.-Y.}\ \bibnamefont
  {Zheng}}, \bibinfo {author} {\bibfnamefont {V.}~\bibnamefont {Achilleos}},
  \bibinfo {author} {\bibfnamefont {Z.-G.}\ \bibnamefont {Chen}}, \bibinfo
  {author} {\bibfnamefont {O.}~\bibnamefont {Richoux}}, \bibinfo {author}
  {\bibfnamefont {G.}~\bibnamefont {Theocharis}}, \bibinfo {author}
  {\bibfnamefont {Y.}~\bibnamefont {Wu}}, \bibinfo {author} {\bibfnamefont
  {J.}~\bibnamefont {Mei}}, \bibinfo {author} {\bibfnamefont {S.}~\bibnamefont
  {Felix}}, \bibinfo {author} {\bibfnamefont {V.}~\bibnamefont {Tournat}}, \
  and\ \bibinfo {author} {\bibfnamefont {V.}~\bibnamefont {Pagneux}},\
  }\href@noop {} {\bibfield  {journal} {\bibinfo  {journal} {New Journal of
  Physics}\ }\textbf {\bibinfo {volume} {22}},\ \bibinfo {pages} {013029}
  (\bibinfo {year} {2020})}\BibitemShut {NoStop}%
\bibitem [{\citenamefont {Coutant}\ \emph {et~al.}(2020)\citenamefont
  {Coutant}, \citenamefont {Achilleos}, \citenamefont {Richoux}, \citenamefont
  {Theocharis},\ and\ \citenamefont {Pagneux}}]{Coutant20}%
  \BibitemOpen
  \bibfield  {author} {\bibinfo {author} {\bibfnamefont {A.}~\bibnamefont
  {Coutant}}, \bibinfo {author} {\bibfnamefont {V.}~\bibnamefont {Achilleos}},
  \bibinfo {author} {\bibfnamefont {O.}~\bibnamefont {Richoux}}, \bibinfo
  {author} {\bibfnamefont {G.}~\bibnamefont {Theocharis}}, \ and\ \bibinfo
  {author} {\bibfnamefont {V.}~\bibnamefont {Pagneux}},\ }\href@noop {}
  {\bibfield  {journal} {\bibinfo  {journal} {Phys. Rev.}\ }\textbf {\bibinfo
  {volume} {B 102}},\ \bibinfo {pages} {214204} (\bibinfo {year} {2020})},\
  \Eprint {http://arxiv.org/abs/2007.13217} {arXiv:2007.13217
  [cond-mat.mes-hall]} \BibitemShut {NoStop}%
\bibitem [{\citenamefont {Coutant}\ \emph {et~al.}(2021)\citenamefont
  {Coutant}, \citenamefont {Achilleos}, \citenamefont {Richoux}, \citenamefont
  {Theocharis},\ and\ \citenamefont {Pagneux}}]{Coutant20b}%
  \BibitemOpen
  \bibfield  {author} {\bibinfo {author} {\bibfnamefont {A.}~\bibnamefont
  {Coutant}}, \bibinfo {author} {\bibfnamefont {V.}~\bibnamefont {Achilleos}},
  \bibinfo {author} {\bibfnamefont {O.}~\bibnamefont {Richoux}}, \bibinfo
  {author} {\bibfnamefont {G.}~\bibnamefont {Theocharis}}, \ and\ \bibinfo
  {author} {\bibfnamefont {V.}~\bibnamefont {Pagneux}},\ }\href@noop {}
  {\bibfield  {journal} {\bibinfo  {journal} {Journal of Applied Physics
  (Acoustic Metamaterials 2021 special issue)}\ }\textbf {\bibinfo {volume}
  {129}},\ \bibinfo {pages} {125108} (\bibinfo {year} {2021})},\ \Eprint
  {http://arxiv.org/abs/2012.15168} {arXiv:2012.15168 [cond-mat.mes-hall]}
  \BibitemShut {NoStop}%
\bibitem [{\citenamefont {Delplace}\ \emph {et~al.}(2011)\citenamefont
  {Delplace}, \citenamefont {Ullmo},\ and\ \citenamefont
  {Montambaux}}]{Delplace11}%
  \BibitemOpen
  \bibfield  {author} {\bibinfo {author} {\bibfnamefont {P.}~\bibnamefont
  {Delplace}}, \bibinfo {author} {\bibfnamefont {D.}~\bibnamefont {Ullmo}}, \
  and\ \bibinfo {author} {\bibfnamefont {G.}~\bibnamefont {Montambaux}},\
  }\href@noop {} {\bibfield  {journal} {\bibinfo  {journal} {Phys. Rev.}\
  }\textbf {\bibinfo {volume} {84}},\ \bibinfo {pages} {195452} (\bibinfo
  {year} {2011})}\BibitemShut {NoStop}%
\bibitem [{\citenamefont {Prodan}\ and\ \citenamefont
  {Schulz-Baldes}(2016)}]{Prodan16}%
  \BibitemOpen
  \bibfield  {author} {\bibinfo {author} {\bibfnamefont {E.}~\bibnamefont
  {Prodan}}\ and\ \bibinfo {author} {\bibfnamefont {H.}~\bibnamefont
  {Schulz-Baldes}},\ }\enquote {\bibinfo {title} {Bulk and boundary invariants
  for complex topological insulators},}\ \ (\bibinfo  {publisher} {Springer},\
  \bibinfo {year} {2016})\ p.~\bibinfo {pages} {40},\ \Eprint
  {http://arxiv.org/abs/1510.08744} {arXiv:1510.08744 [math-ph]} \BibitemShut
  {NoStop}%
\bibitem [{\citenamefont {Kariyado}\ and\ \citenamefont
  {Hu}(2017)}]{Kariyado17}%
  \BibitemOpen
  \bibfield  {author} {\bibinfo {author} {\bibfnamefont {T.}~\bibnamefont
  {Kariyado}}\ and\ \bibinfo {author} {\bibfnamefont {X.}~\bibnamefont {Hu}},\
  }\href@noop {} {\bibfield  {journal} {\bibinfo  {journal} {Scientific
  reports}\ }\textbf {\bibinfo {volume} {7}},\ \bibinfo {pages} {1} (\bibinfo
  {year} {2017})}\BibitemShut {NoStop}%
\bibitem [{\citenamefont {Inui}\ \emph {et~al.}(1994)\citenamefont {Inui},
  \citenamefont {Trugman},\ and\ \citenamefont {Abrahams}}]{Inui94}%
  \BibitemOpen
  \bibfield  {author} {\bibinfo {author} {\bibfnamefont {M.}~\bibnamefont
  {Inui}}, \bibinfo {author} {\bibfnamefont {S.}~\bibnamefont {Trugman}}, \
  and\ \bibinfo {author} {\bibfnamefont {E.}~\bibnamefont {Abrahams}},\
  }\href@noop {} {\bibfield  {journal} {\bibinfo  {journal} {Phys. Rev.}\
  }\textbf {\bibinfo {volume} {B 49}},\ \bibinfo {pages} {3190} (\bibinfo
  {year} {1994})}\BibitemShut {NoStop}%
\bibitem [{\citenamefont {Mondragon-Shem}\ \emph {et~al.}(2014)\citenamefont
  {Mondragon-Shem}, \citenamefont {Hughes}, \citenamefont {Song},\ and\
  \citenamefont {Prodan}}]{Mondragon14}%
  \BibitemOpen
  \bibfield  {author} {\bibinfo {author} {\bibfnamefont {I.}~\bibnamefont
  {Mondragon-Shem}}, \bibinfo {author} {\bibfnamefont {T.~L.}\ \bibnamefont
  {Hughes}}, \bibinfo {author} {\bibfnamefont {J.}~\bibnamefont {Song}}, \ and\
  \bibinfo {author} {\bibfnamefont {E.}~\bibnamefont {Prodan}},\ }\href@noop {}
  {\bibfield  {journal} {\bibinfo  {journal} {Phys. Rev. Lett.}\ }\textbf
  {\bibinfo {volume} {113}},\ \bibinfo {pages} {046802} (\bibinfo {year}
  {2014})}\BibitemShut {NoStop}%
\end{thebibliography}%

\end{document}